\documentclass[aps,twocolumn,prl,nofootinbib,superscriptaddress,geometry,preprintnumbers]{revtex4-1}

\usepackage{amsmath,amssymb} 
\usepackage{accents}
\usepackage{nccmath}
\usepackage{color}
\usepackage[bookmarks=true,hyperfigures=true]{hyperref}
\usepackage{graphicx}
\usepackage{tensor}
\usepackage{empheq}
\usepackage{nicefrac}
\usepackage{shuffle}
\usepackage{slashed}
\usepackage{tikz-feynman}
\tikzfeynmanset{compat=1.1.0}

\usepackage{color}

\usepackage{feynmp-auto}
\usepackage{feynmp} % we can remove the force option later
\allowdisplaybreaks[3]

\usepackage[font=small,labelfont=bf,width=0.85\textwidth]{caption}

\let\oldbfseries=\bfseries
\let\oldmdseries=\mdseries
\let\oldnormalfont=\normalfont
\renewcommand{\bfseries}{\oldbfseries\boldmath}
\renewcommand{\mdseries}{\oldmdseries\unboldmath}
\renewcommand{\normalfont}{\oldnormalfont\unboldmath}

\makeatletter
\newlength{\apb@width}
\newcommand{\autoparbox}[2][c]{\settowidth{\apb@width}{#2}\parbox[#1]{\apb@width}{#2}}

\makeatother

\newcommand{\nn}{\nonumber}

\newcommand{\remark}[2][.]{{\color{red}\renewcommand{\bfdefault}{b}\rmfamily\if.#1\else\textbf{#1:} \fi#2}}

\newcommand{\be}{\begin{equation}}
\newcommand{\ee}{\end{equation}}
\newcommand{\beq}{\begin{equation}}
\newcommand{\eeq}{\end{equation}}
\newcommand{\bal}[1]{\setlength{\jot}{1em}\begin{align}\begin{aligned}#1\end{aligned}\end{align}}

\newcommand{\bma}{\begin{pmatrix}}
\newcommand{\ema}{\end{pmatrix}}
\newcommand{\ba}{\begin{eqnarray}}
\newcommand{\ea}{\end{eqnarray}}
\newcommand{\tf}{\tilde f}

\ifx\genfrac\sdflkaj\else\fi

\newcommand \widebar [1] {\overline{#1}}

\newcommand{\cA}{\mathcal{A}}

\def\l<{\langle}\def\r>{\rangle}

\makeatletter
\newcommand{\namedref}[2]{\hyperref[#2]{#1~\ref*{#2}}}
\newcommand{\secref}{\@ifstar{\namedref{Section}}{\namedref{sec.}}}
\newcommand{\subsecref}{\@ifstar{\namedref{Subsection}}{\namedref{subsec.}}}
\newcommand{\appref}{\@ifstar{\namedref{Appendix}}{\namedref{app.}}}
\newcommand{\tabref}{\@ifstar{\namedref{Table}}{\namedref{tab.}}}
\newcommand{\figref}{\@ifstar{\namedref{Figure}}{\namedref{fig.}}}
\makeatother

\def\[{\begin{equation}}
\def\]{\end{equation}}
\def\<{\begin{eqnarray}}
\def\>{\end{eqnarray}}

\newcommand{\Tr}{\mathop{\mathrm{Tr}}}

\newcommand{\eqn}[1]{(\ref{#1})}

\def\bea{\begin{align}}
\def\eea{\end{align}}
\def\be{\begin{equation}}
\def\ee{\end{equation}}

\begin{document}
	\newtheorem{theorem}{Theorem}
	\newtheorem{theorem1}{Theorem}
\thispagestyle{empty}

%%%%%%%%%%%%%%%%%%%%%%%%%%%%%%%%%%%%%%%%%%%%%%%%%%%%%%%%%%%%%%%%%%%%%%%%%%%%%%%%

\title{New relations for graviton-matter amplitudes }
%\date{\today}
\author{Jan Plefka}
\affiliation{Institut f\"ur Physik und IRIS Adlershof, Humboldt-Universit\"at zu Berlin, 
  Zum Gro{\ss}en Windkanal 6, 12489 Berlin, Germany}
\affiliation{Theoretical Physics Department, CERN, 1211 Geneva 23, Switzerland}
\author{Wadim Wormsbecher}
\affiliation{Institut f\"ur Physik und IRIS Adlershof, Humboldt-Universit\"at zu Berlin, 
  Zum Gro{\ss}en Windkanal 6, 12489 Berlin, Germany}
\email{jan.plefka@physik.hu-berlin.de, wadim.wormsbecher@physik.hu-berlin.de}
\preprint{HU-EP-18/13, CERN-TH-2018-095}

\begin{abstract}
We present new relations for scattering amplitudes of color ordered 
gluons, massive quarks and  scalars minimally coupled to gravity. 
Tree-level amplitudes of arbitrary matter and gluon multiplicities 
involving one graviton are reduced to partial amplitudes in QCD or scalar QCD. The obtained relations are a direct
generalization of the recently found Einstein-Yang-Mills relations. 
The proof of the new relation  employs a simple diagrammatic
argument trading the graviton-matter couplings to an `upgrade' of a gluon coupling with a color-kinematic replacement rule enforced. The use of the
Melia-Johansson-Ochirov color basis is a key element of the reduction.
We comment on the generalization to multiple gravitons in the single color trace case.
\end{abstract}

\maketitle

\section{Introduction}

At the Lagrangian level 
Einstein's theory of gravity and Yang-Mills (YM) gauge theories 
look quite different.  Nonetheless, in a perturbative quantization on a flat space-time background intimate relations between their S-matrices exist, allowing one to express pure graviton scattering amplitudes through pure gluon scattering data.
The  first such connection are the Kawai-Lewellen-Tye relations 
\cite{Kawai:1985xq}
derived from the string theory.
%theoretic origin of the tree-level field-theory S-matrices. 
%Some  time ago 
Later Bern, Carrasco and Johansson (BCJ) \cite{Bern:2008qj, Bern:2010yg, Bern:2010ue} introduced a double-copy construction for graviton amplitudes from gluons.
Here Lie-algebra like relations for the kinematic building blocks 
of gluon amplitudes were identified.
The double copy technique 
may be  used  to generate loop-level integrands of gravitational theories from the simpler
 gauge-theory ones, this being the state-of-the-art method for higher loop computations in
 (super)-gravity.
% , see for instance \cite{Bern:2012uf, Bern:2012cd, Bern:2013uka, Bern:2014sna,Bern:2017ucb}. 
For the phenomenologically most relevant case of gravity minimally coupled to generic
non-abelian gauge and matter fields of spins 1, $\nicefrac{1}{2}$ and 0, our knowledge is
less complete. Early results for  maximally-helicity violating (MHV) tree amplitudes 
\cite{Selivanov:1997aq,Selivanov:1997ts,Bern:1999bx} were rather recently dramatically
extended to the tree-level sector of Einstein-Yang-Mills (EYM) amplitudes. Here 
the full modern arsenal of amplitude techniques was employed, starting from the field theory
limit of string amplitudes \cite{Stieberger:2016lng}, field theoretical considerations
\cite{Nandan:2016pya,delaCruz:2016gnm} using the Cachazo-He-Yuan (CHY) formalism \cite{Cachazo:2013gna, Cachazo:2014nsa, Cachazo:2014xea}, as well as double copy
methods \cite{Chiodaroli:2017ngp}. In fact, 
the complete reduction of the EYM tree-level S-matrix to the
YM one was accomplished in \cite{Fu:2017uzt,Teng:2017tbo,Du:2017gnh}.
This in combination with the
existing result for all tree-level color-ordered gluon amplitudes \cite{Drummond:2008cr,Dixon:2010ik,Bourjaily:2010wh}, constitutes the complete solution for the EYM $S$-matrix at tree level. First results for pure EYM amplitudes at one-loop level at multiplicity four were recently reported in
\cite{Nandan:2018ody}.

In this letter we report on an extension of these results to generic massive non-abelian
matter fields, i.e.~QCD and scalar QCD with $N_{f}$ flavors
minimally coupled to Einstein's gravity. Again, we are able
to present compact formulae to express single graviton-quark-gluon and graviton-scalar amplitudes
in linear combinations of non-gravitational amplitudes. We comment on generalizations
to higher graviton multiplicity in the discussion.

\section{Einstein-Yang Mills}
 
%To begin with let us quickly rederive the  emission of a single graviton from an $n$-gluon scattering.
Gluon amplitudes may be color decomposed in various
bases. A particular useful  and minimal one, as it generalizes to QCD, is the Del Duca-Dixon-Maltoni (DDM) basis \cite{DelDuca:1999rs} which
organizes the $n$-gluon scattering amplitude in a basis of $(n-2)!$ partial amplitudes
\begin{align}
\cA^{\text{tree}}_{n} =  &\sum_{\sigma \in S_{n-2}(\{3,\ldots, n\})}
 C(1,2,\sigma)\,  A^{\text{YM}}(1,2,\sigma)\, .
\end{align}
with the DDM factors
\be\nn
C(1,2,\sigma_{3},\ldots,\sigma_{n}):=
\tf^{a_{2}a_{\sigma(3)}b_{1}} \tf^{b_{1}a_{\sigma(4)}b_{2}} \ldots
\tf^{b_{n-3}a_{\sigma(n)}a_{1}}
\ee
where $\tf^{abc}=i\,f^{abc}$ are the structure constants of the gauge group (our conventions are summarized in the appendix).
Clearly, also for an EYM amplitude involving a single graviton and at leading order in gravitational coupling $\kappa$ (implying a color single trace structure) an identical color decomposition
in the DDM basis  may be applied 
\begin{align}
\cA^{\text{tree}}_{n;1} =  &\sum_{\sigma \in S_{n-2}(\{3,\ldots, n\})}
C(1,2,\sigma)\, A^{\text{EYM}}(1,2,\sigma;p)\, ,
\end{align}
where the graviton leg $p$ is of course not participating in the color ordering.
In \cite{Stieberger:2016lng} an intriguingly simple representation of this partial
EYM-amplitude was derived from a string theory consideration
\begin{align}
\label{ST-relation}
A^{\text{EYM}}(1,2,\ldots,n;p) =& \\
\frac{\kappa}{2g}\sum_{i=2}^{n} (\varepsilon_{p}\cdot X_{i})\, &
A^{\text{YM}}(1,2,\ldots,i,p,i+1,\ldots)\, . \nn
\end{align}
Here the graviton polarization is written as $\varepsilon^{\mu\nu}_{p}=\varepsilon_{p}^{\mu}
\varepsilon_{p}^{\nu}$. Moreover $X_{i}=\sum_{j=2}^{i}k_{j}$ denotes the region momentum. In a sense \emph{one half} of the graviton has turned into a gluon evenly distributed amongst the $n$ gluons. \\
Indeed the relation \eqn{ST-relation} immediately follows from the consistency of \emph{soft-limits}. Both gravitons \cite{Weinberg:1964ew} and gluons  \cite{Low:1958sn} (resp.~photons)
obey universal factorization properties in the soft limit $p\to 0$ 
\begin{align}
A^{\text{EYM}}(1,\ldots, n;p) \stackrel{p\to 0}{=} &\sum_{i=1}^{n}
\frac{(\varepsilon_{p}\cdot k_{i})^{2}}{k_{i}\cdot p}\, A^{\text{YM}}(1,\ldots, n)  \nn \\
A^{\text{YM}}(\ldots,i,p,i+1,\ldots) \stackrel{p\to 0}{=} & \label{Soft-limits}
\\
\Bigl( \frac{\varepsilon_{p}\cdot k_{i}}{k_{i}\cdot p}& -
 \frac{\varepsilon_{p}\cdot k_{i+1}}{k_{i+1}\cdot p}\Bigr)  \, A^{\text{YM}}(1,\ldots, n) 
 \, .
 \nn
\end{align}
Our convention is that all external momenta are incoming. If one starts with \eqn{ST-relation} as an ansatz with an  a priori undetermined $X_{p}$ one
quickly arrives at the consistency conditions
\be
X_{i}-X_{i-1}=k_{i} \quad , \quad X_{2}=k_{2} \quad \text{and} \quad
X_{n}=-k_{1}
\ee
upon taking $p$ soft. These relations are solved for the region momenta $X_{i}=\sum_{j=2}^{i}k_{j}$.
In addition, the second consistency requirement of \emph{gauge invariance} of \eqn{ST-relation},
$\varepsilon_{p}^{\mu}\to p^{\medskip	\mu}$, immediately yields the famous BCJ relation \cite{Bern:2008qj}
\be
0=\sum_{i=2}^{n} (p\cdot X_{i})\, 
A^{\text{YM}}(1,2,\ldots,i,p,i+1,\ldots)\, ,
\label{BCJ-relations}
\ee
an essential ingredient of
the double-copy construction \cite{Bern:2010yg, Bern:2010ue}. So indeed
\eqn{ST-relation} is entirely constrained by soft limits and gauge invariance.

\section{Einstein-QCD}

We now generalize \eqn{ST-relation} to QCD minimally coupled to gravity (EQCD). 
%The results carry over to scalar QCD as we will discuss later. 
For this we
need to first discuss the issue of color ordering in the presence of
quarks and anti-quarks in the fundamental representation 
$T^{a}_{i\bar{j}}$ of the gauge group. A very useful color basis for this
was provided by Melia \cite{Melia:2013epa} and refined by Johansson-Ochirov 
\cite{Johansson:2015oia}, which we term the MJO-basis.
% and briefly review. 
An $n$-particle QCD amplitude consists of $k$ quark-anti-quark pairs and $n-2k$
gluons.  Without loss of generality we take the flavors of \emph{all} $k$ quark lines 
to be distinct. The primitive amplitudes in the MJO-basis are given by
\be
\Bigl \{ A(\underbar{1},\bar{2},\sigma) \Bigr | \sigma\in \text{Dyck}_{k-1} 
\times \{\text{gluon insertions}\}_{n-2k}
\Bigr \} \, . \nn
\ee
Quarks and anti-quarks are marked with under-scores or over-scores, respectively.
In the partial QCD amplitude $A(\underbar{1},\bar{2},\sigma) $ the permutation of the
last $(n-2)$ arguments must form a  Dyck word. The most intuitive defintion is that a Dyck word corresponds
to well formed bracket expressions with quarks preceded by opening and antiquarks followed by closing
brackets. 
%A pure quark $n=6$, $k=3$ example with quark flavor lines $\underbar{3}\leftarrow\bar{4}$
%and $\underbar{5}\leftarrow\bar{6}$ would read
%%
%\begin{align}
%& (\underbar{3},\bar{4},\underbar{5},\bar{6}), (\underbar{5},\bar{6},\underbar{3},\bar{4}),
%(\underbar{3},\underbar{5},\bar{6},\bar{4}),(\underbar{5},\underbar{3},\bar{4},\bar{6}) \\
%\Leftrightarrow\quad & \{3 4\}\{56\},\, \{56\}\{34\} ,\, \{3 \{ 5 6\} 4\} ,\, 
%\{5\{ 3 4\}6\} \, .\nn
%\end{align}
%%
In the MJO-basis a QCD amplitude may then be decomposed as
\be
\cA_{n,k}^{\text{tree}}= \sum_{\sigma\in\text{MJO basis}}^{\chi(n,k)}\,
C(\underbar{1},\bar{2},\sigma)\, A^{\text{QCD}}(\underbar{1},\bar{2},\sigma)
\ee
where $\chi(n,k)=\frac{(n-2)!}{k!}$  is the dimension of the
basis \cite{Johansson:2015oia}. Using the bracket notation
the color factors are given
by
\be
\label{CC-def}
\left.
C(\underbar{1},\bar{2},\sigma) = (-)^{k-1}\, \{2|\sigma|1\}
\phantom{\begin{matrix}
{}_{q }\\
{}_{q}\\ 
{}_{g}\\
\end{matrix}
}\right|
\begin{matrix}
\hspace{0.05cm}q\to \{q|\, T^{b}\otimes \Xi^{b}_{l-1} \\
\bar{q}\to |q\}~~~~~~~~~~~~~ \\ 
g \to \Xi^{a_{g}}_{l}~~~~~~~~~~~~\\
\end{matrix}
\ee
Here a level of `nestedness' $l$ has been introduced, it is the number of anti-quarks minus the number of quarks to the left of the position in the Dyck word, and reflected in the tensor product structure.
The important object $\Xi^{a}_{l}$ takes the form
\be
\Xi_{l}^{a}=\sum_{s=1}^{l}\underbrace{1\otimes \ldots \otimes 1\otimes \overbrace{T^{a}\otimes 1\otimes
\ldots 1\otimes \widebar{1}}^s}_{l}\, .
\label{xidef}
\ee
The $\Xi^{a}_{l}$ form a representation of the gauge group Lie algebra $[\Xi^{a}_{l},\Xi^{b}_{l}]=
\tf^{abc}\Xi^{c}_{l}$.
The explicit $C(\underbar{1},\bar{2},\sigma)$'s are sums of 
 products of $k$ strings of generators $(T^{a_{1}}\ldots T^{a_{r}})_{i\bar j}$ contracted over adjoint indices 
 
Clearly then, an Einstein-QCD amplitude at leading order in $\kappa$ (single color trace) involving a single-graviton enjoys  a color
decomposition in the MJO basis 
\be
\label{EQCD-full}
\cA_{n,k;1}^{\text{tree}}= \sum_{\sigma\in\text{MJO basis}}^{\chi(n,k)}\,
C(\underbar{1},\bar{2},\sigma)\, A^{\text{EQCD}}(\underbar{1},\bar{2},\sigma;p) \, .
\ee
The central result of this letter is, that this partial EQCD-amplitude takes the form
\begin{align}
\label{EQCD-claim}
A^{\text{EQCD}}(1,2,\ldots,n;p) =& \\
\frac{\kappa}{2g}\sum_{i=2}^{n} (\varepsilon_{p}\cdot X_{i})\, &
A^{\text{QCD}}(1,2,\ldots,i,p,i+1,\ldots)\, . \nn
\end{align}
in complete analogy to \eqn{ST-relation}. A gauge transformation on leg $p$ yields
the BCJ relations \eqn{BCJ-relations} in QCD \cite{Johansson:2015oia}
which were proven in \cite{delaCruz:2015dpa}. 
The relation \eqn{EQCD-claim} is also consistent with the soft limits as the soft behavior of gravitons and gluons is \emph{universal}, 
i.e.~\eqn{Soft-limits} also holds for $\{1,\ldots,n\}$ being (anti)-quarks \emph{or} gluons. 
These consistencies are strong evidence for the correctness of \eqn{EQCD-claim}.
\section{Diagrammatic Proof}

In order to prove \eqn{EQCD-claim} consider the Feynman diagrammatic representation
of the color dressed $\cA^{\text{tree}}_{n,k;1}$ of \eqn{EQCD-full}. Starting from a pure QCD color dressed $n$-point amplitude
$\cA^{\text{tree}}_{n,k}$ the EQCD-amplitude may be obtained by attaching the graviton leg 
with data $\{\varepsilon_{p}^{\mu}\varepsilon_{p}^{\nu},p^{\rho}\}$ to \emph{all} propagators and vertices of the
lower point QCD amplitude. The relevant vertices are depicted in FIG.~\ref{fig:EQCD-rules}, their
mathematical expressions \cite{TheoThesis} are collected in the appendix.
\begin{figure}
 \includegraphics[width=1.2cm]{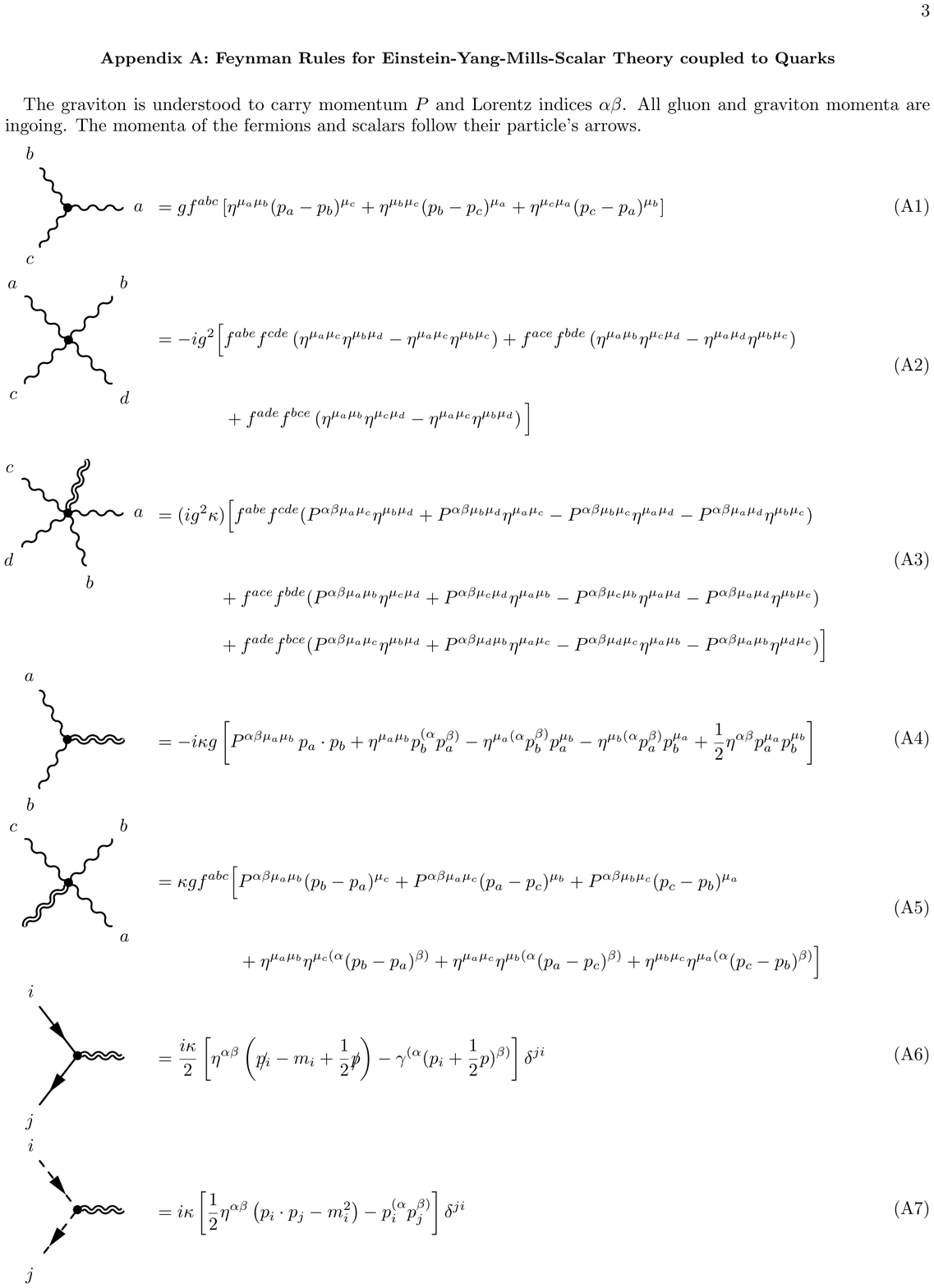}\quad
 \raisebox{0.15cm}{
  \includegraphics[width=1.2cm]{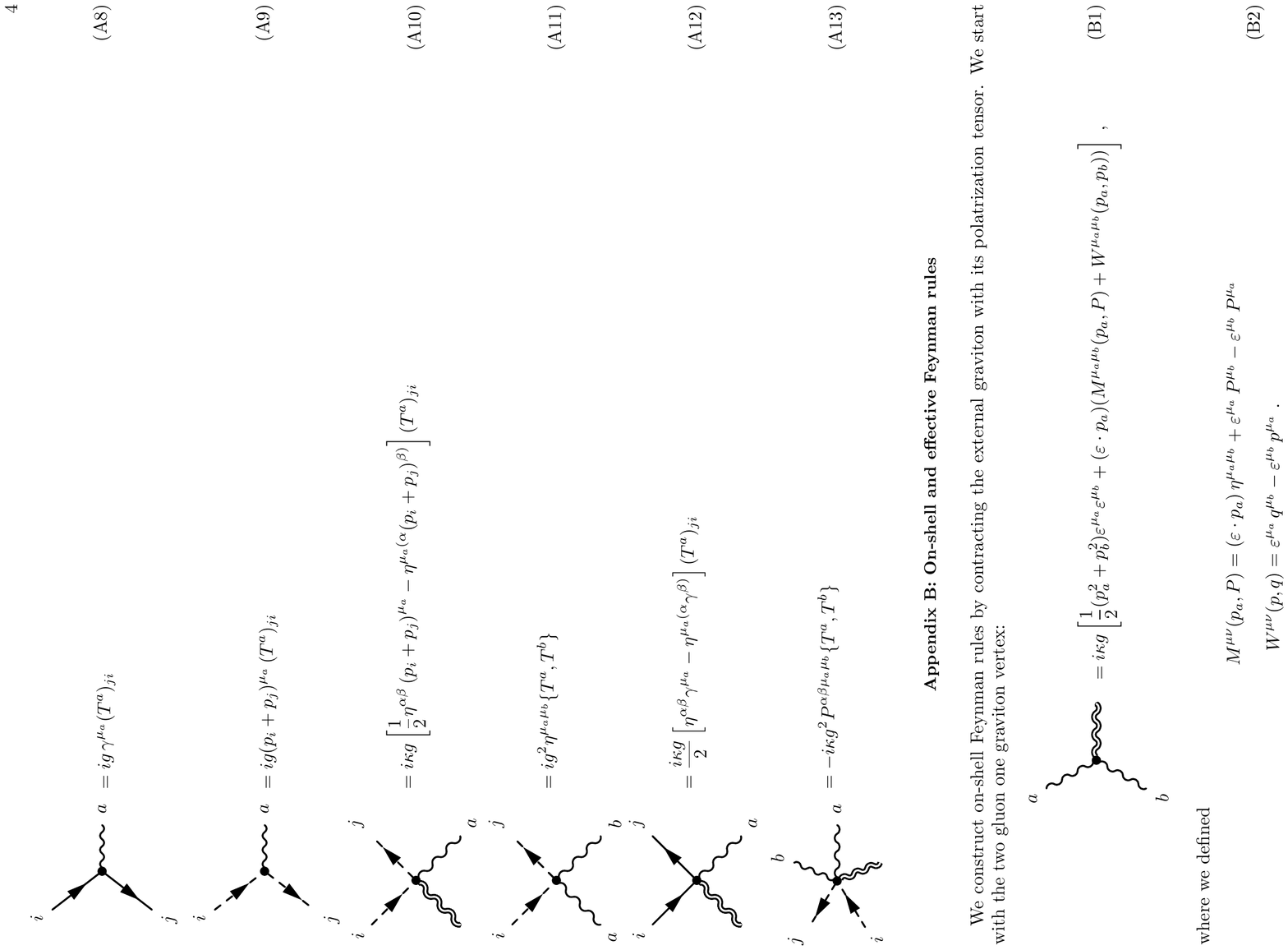}}\quad
  \includegraphics[width=1.2cm]{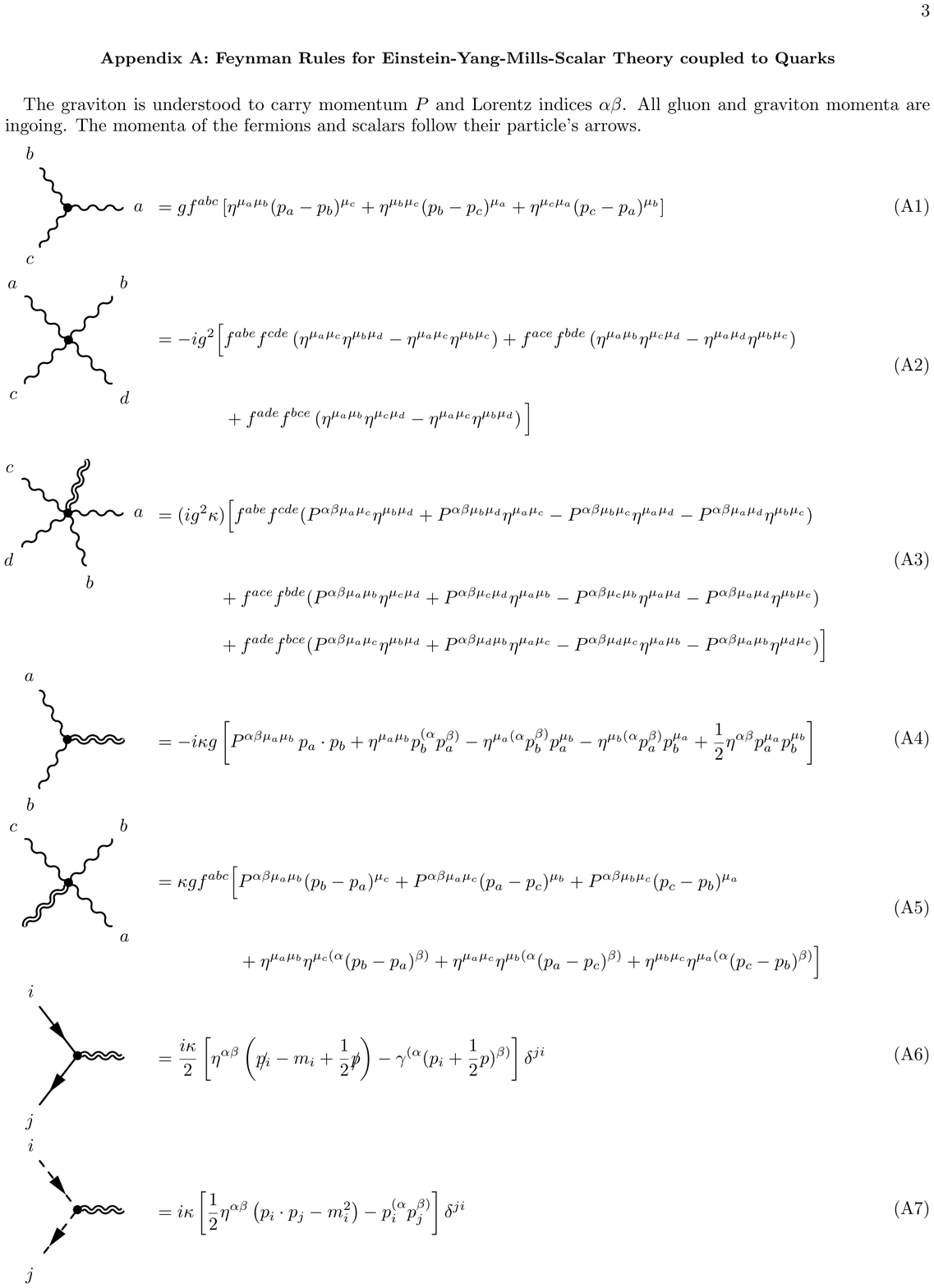}\quad
    \includegraphics[width=1.2cm]{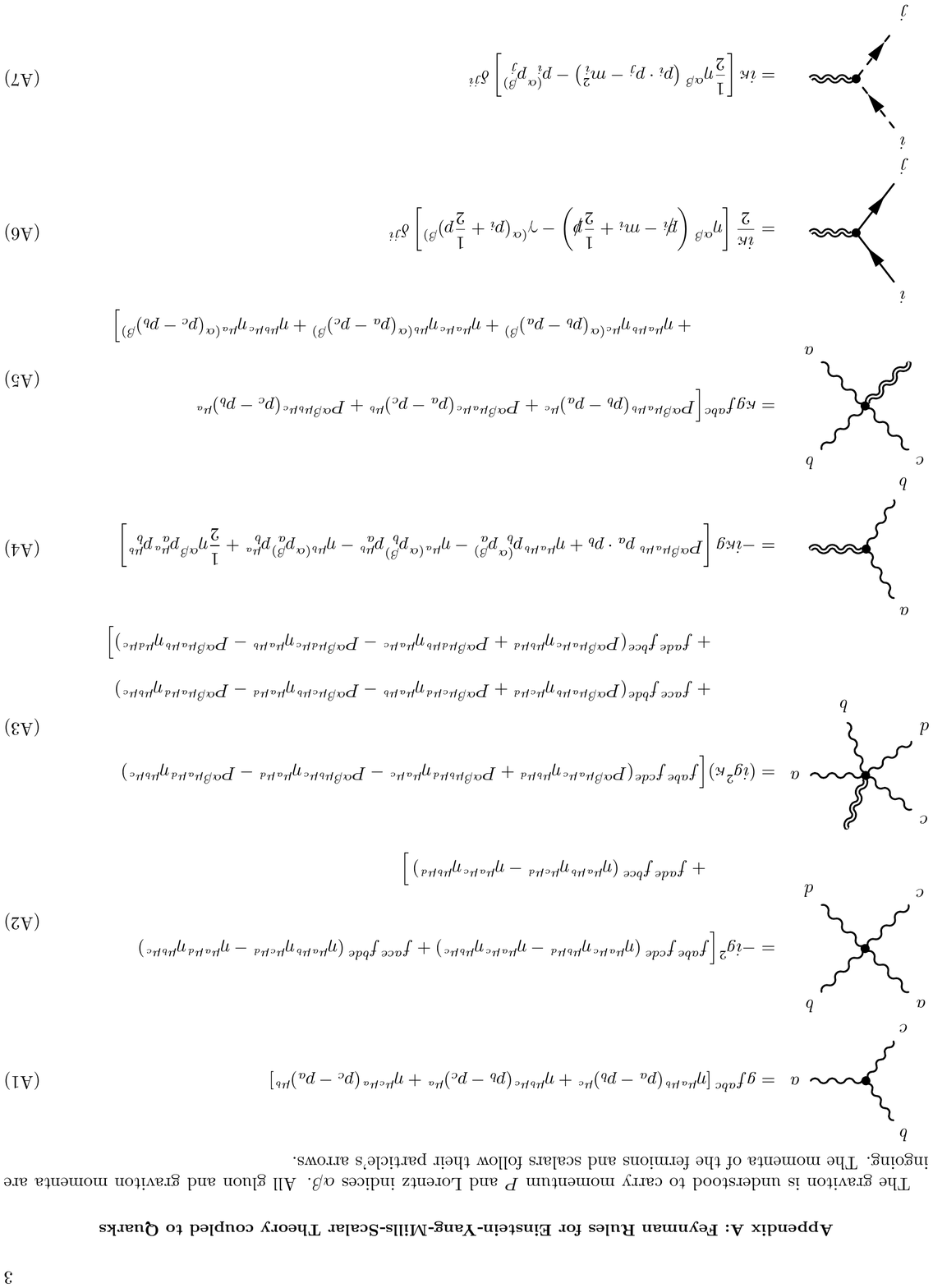}\quad
      \includegraphics[width=1.2cm]{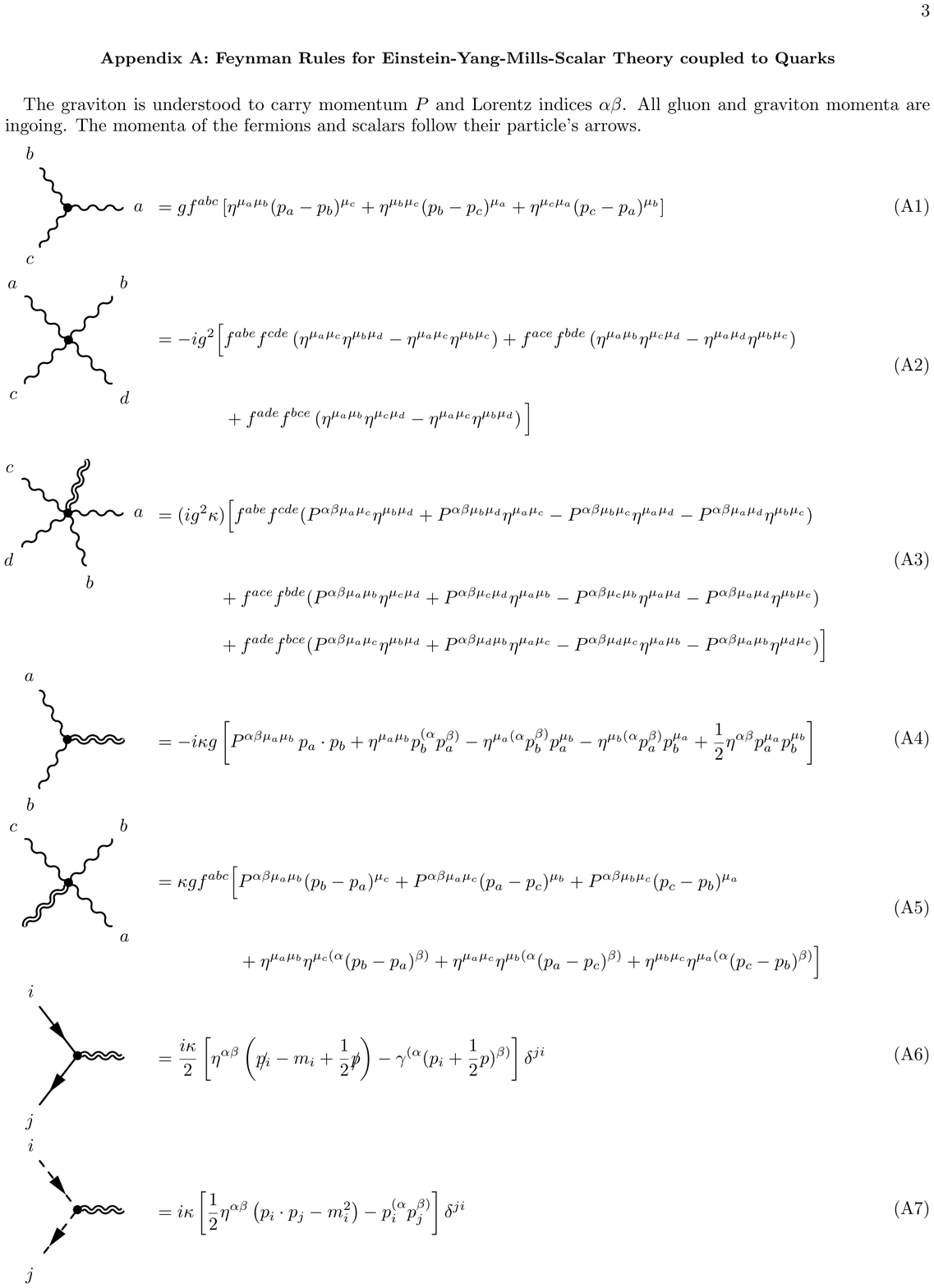}
 \caption{\label{fig:EQCD-rules} Single graviton Einstein-QCD vertices.}
  \end{figure}
Putting the graviton leg on-shell results in a subtle simplification. Let us look at the pure gluon-graviton vertices
first. Here one observes a decomposition as
$$
\includegraphics[height=1.8cm]{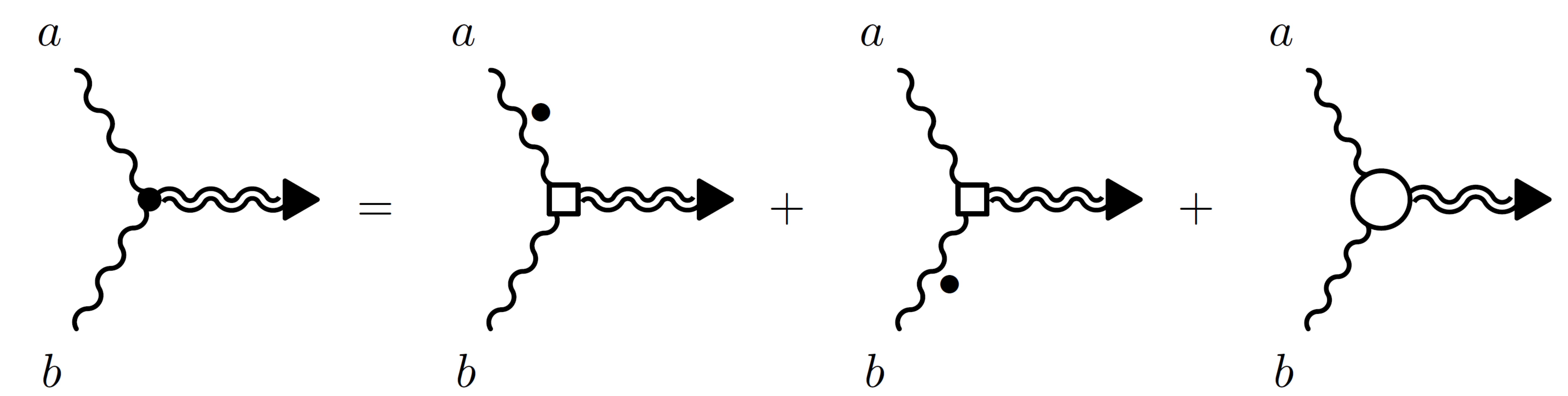}\quad\raisebox{0.85cm}{.}
$$
where the triangle attached to a leg indicates on-shellness. The dot on top of the gluon line denotes the appearance of the inverse massless propagator $ k_{a}^{2}$ or $k_{b}^{2}$ which would
be attached on the corresponding leg. Interestingly, the last term above 
represents an effective EYM vertex intimately related to the
pure YM three-gluon vertex through a color-kinematical replacement rule 
\be
\label{CCid1}
\left.
\raisebox{-0.83cm}{\includegraphics[height=1.8cm]{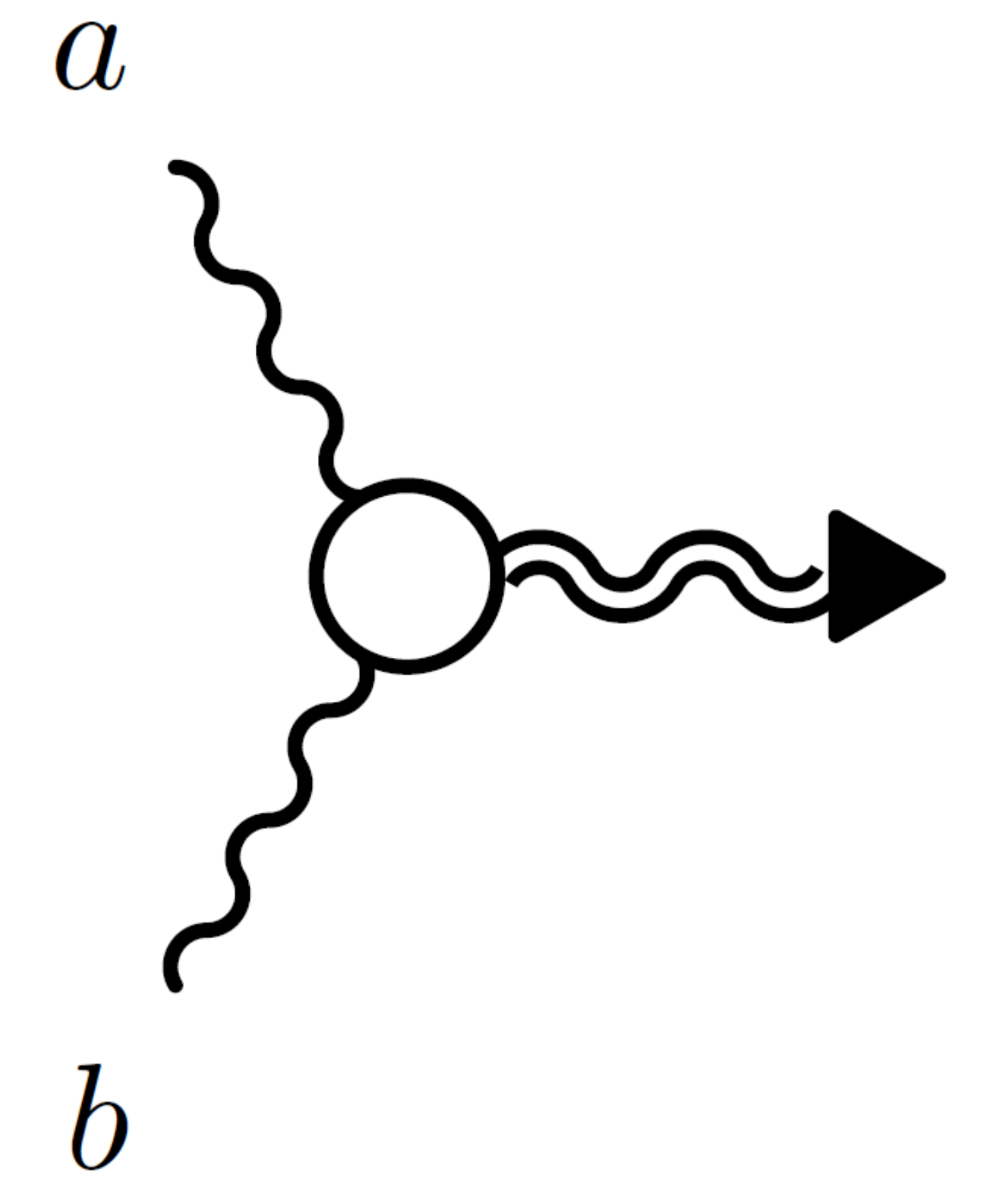}}  = 
\raisebox{-0.83cm}{\includegraphics[height=1.8cm]{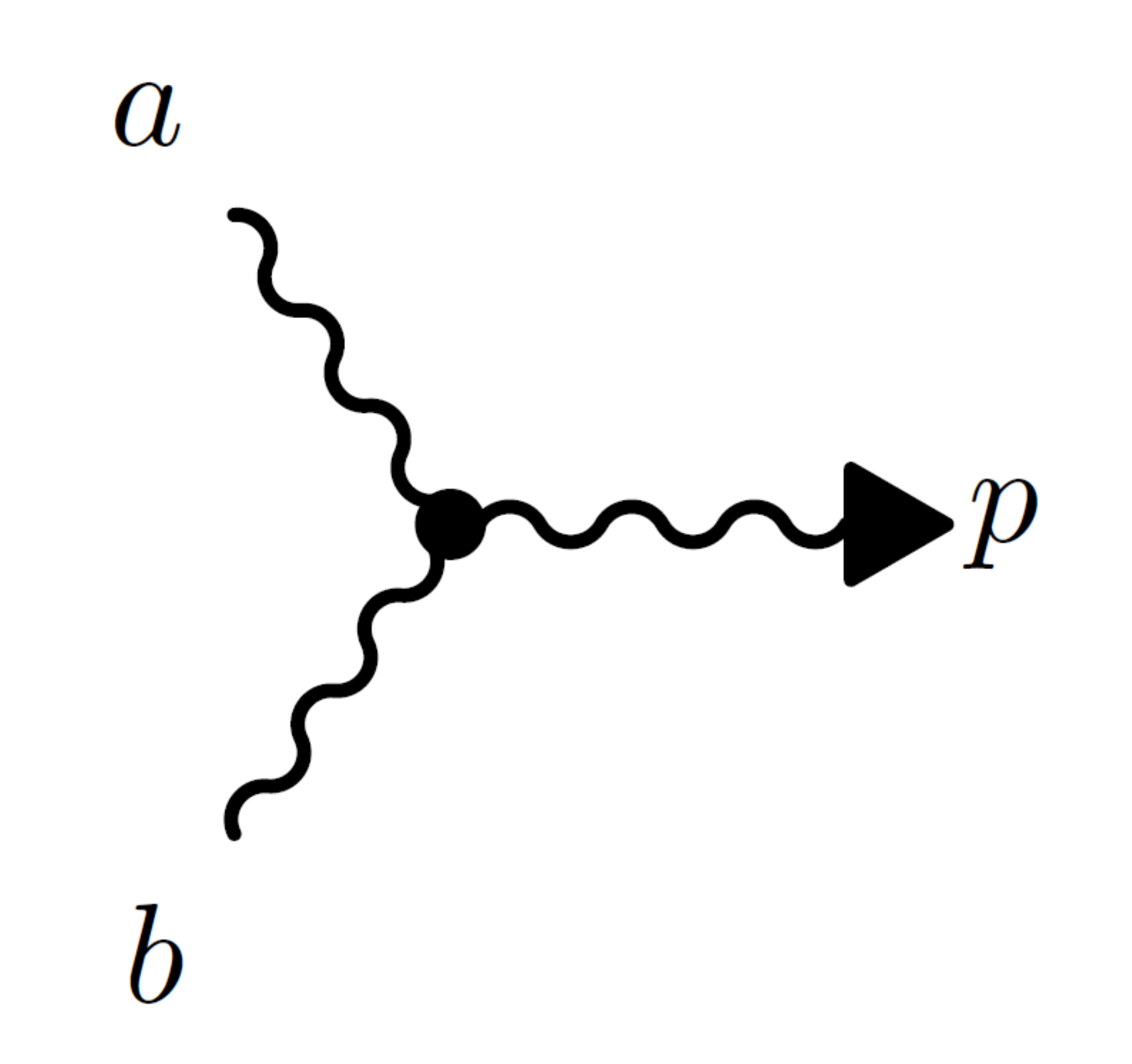}}\right|\raisebox{-0.5cm}{$f^{apb}\to -i\frac{\kappa}{2\,g}\,(\varepsilon_{p}\cdot k_{a})\, \delta_{ab}$}~,
\ee
which we demonstrate in the appendix.
In this relation one replaces the color factor $f^{abc}$ by a kinematic expression assigning a momentum to every leg in color space. The convention is that all momenta are inflowing. 
In this argument the Ward identity has been used for
the gluon legs $a$ and $b$ as they attach to invariant subamplitudes.
Similarly the graviton-three-gluon vertex decomposes as
$$
\includegraphics[height=1.8cm]{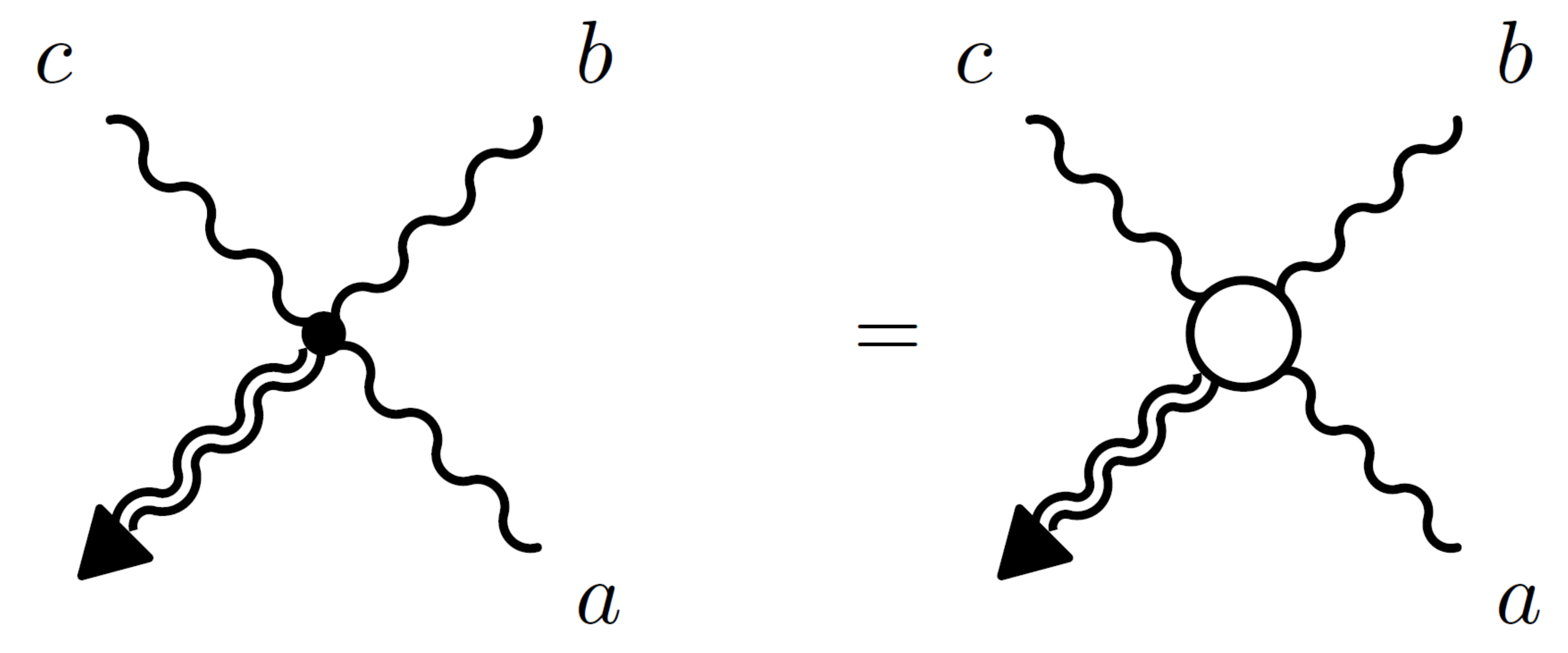} \raisebox{0.8cm}{$- \sum\limits_{\text{cyclic}(abc)}$} \includegraphics[height=1.8cm]{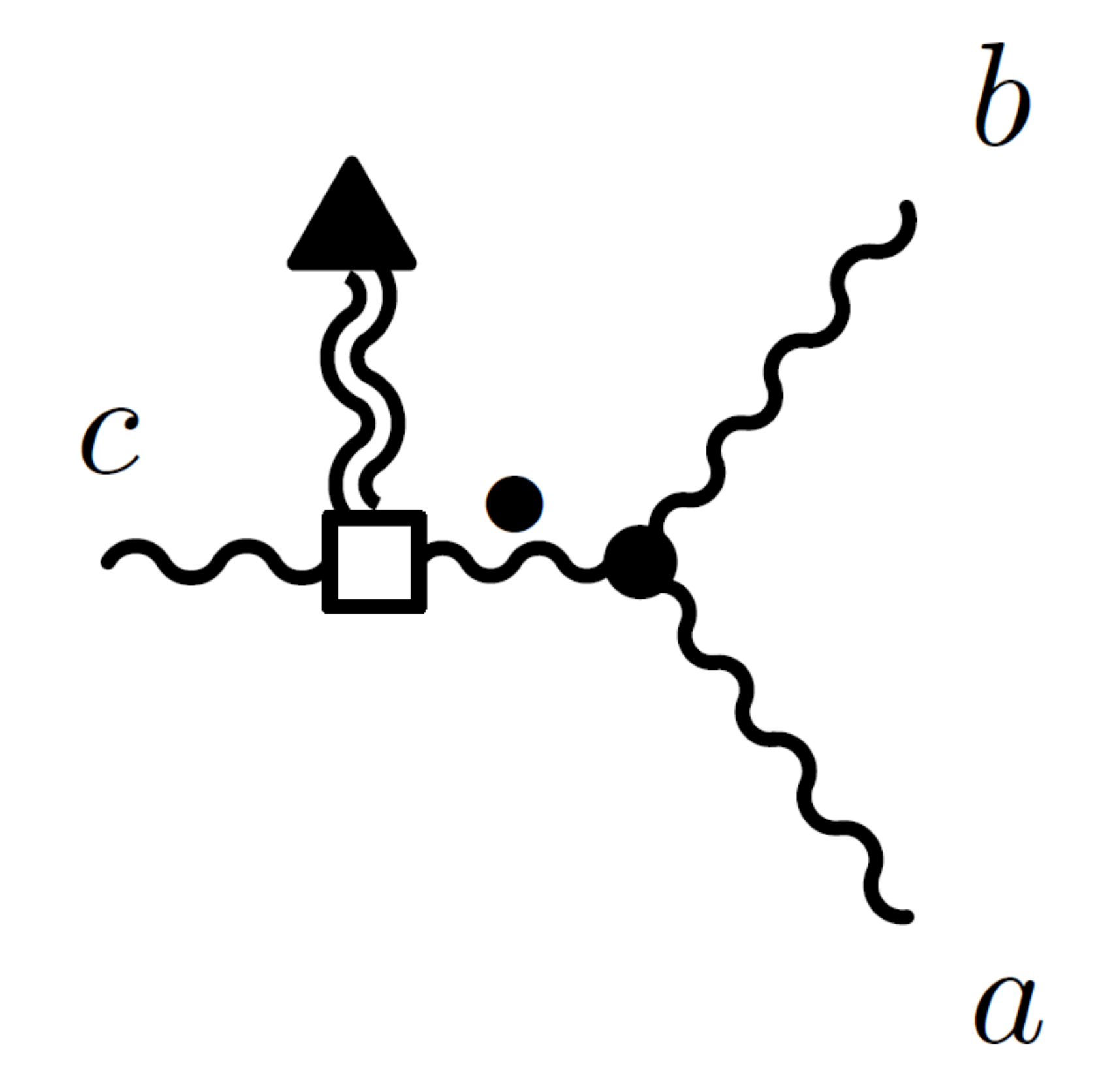} \, \raisebox{0.7cm}{,}
$$
where again the dots on the top of the gluon lines denote inverse gluon propagators.
Just as before the effective EYM vertex follows from the four gluon vertex upon color-kinematic
replacement
\be
\label{CCid2}
\left.
\raisebox{-0.83cm}{\includegraphics[height=1.8cm]{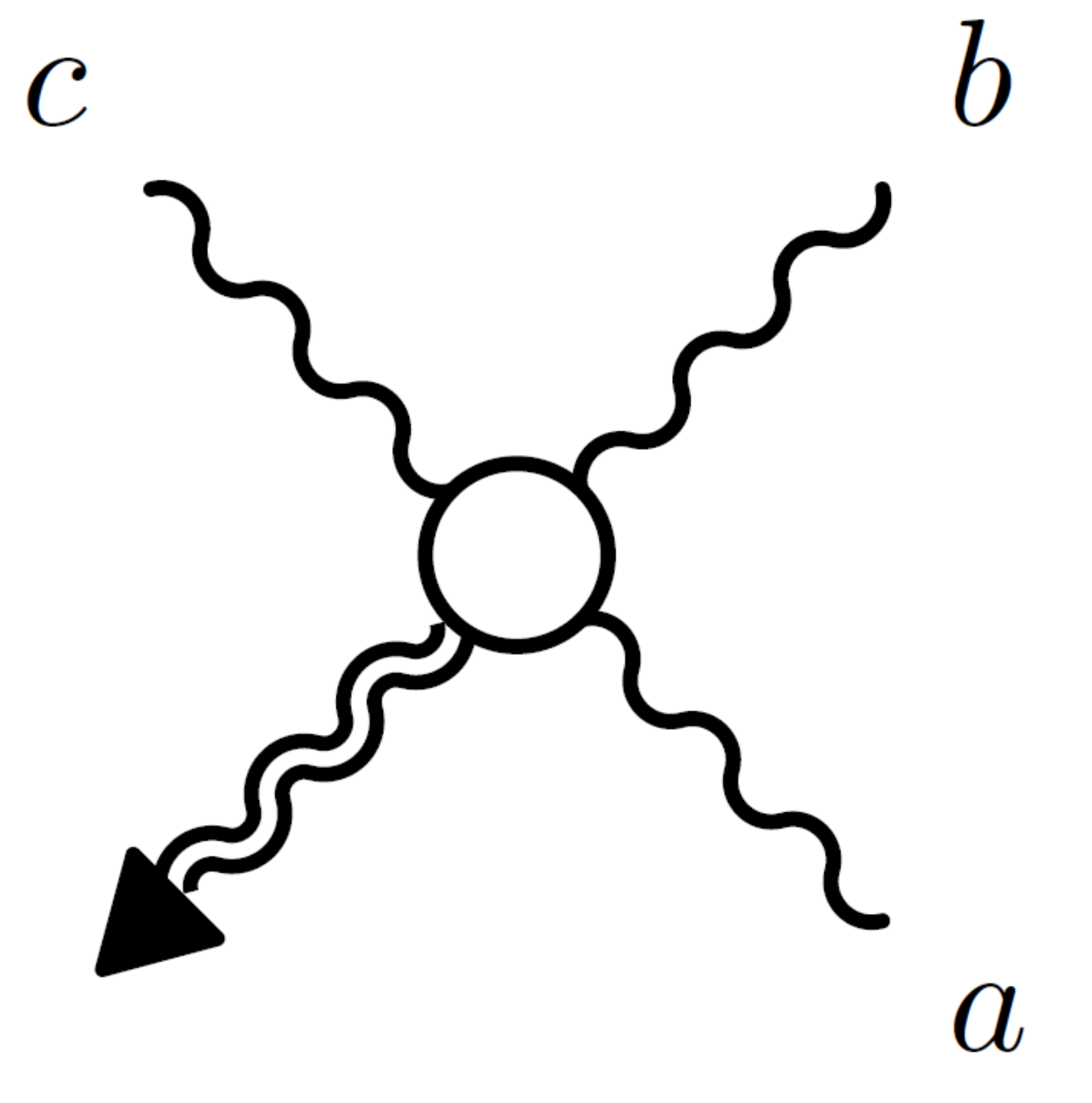}}  = 
\raisebox{-0.83cm}{\includegraphics[height=1.8cm]{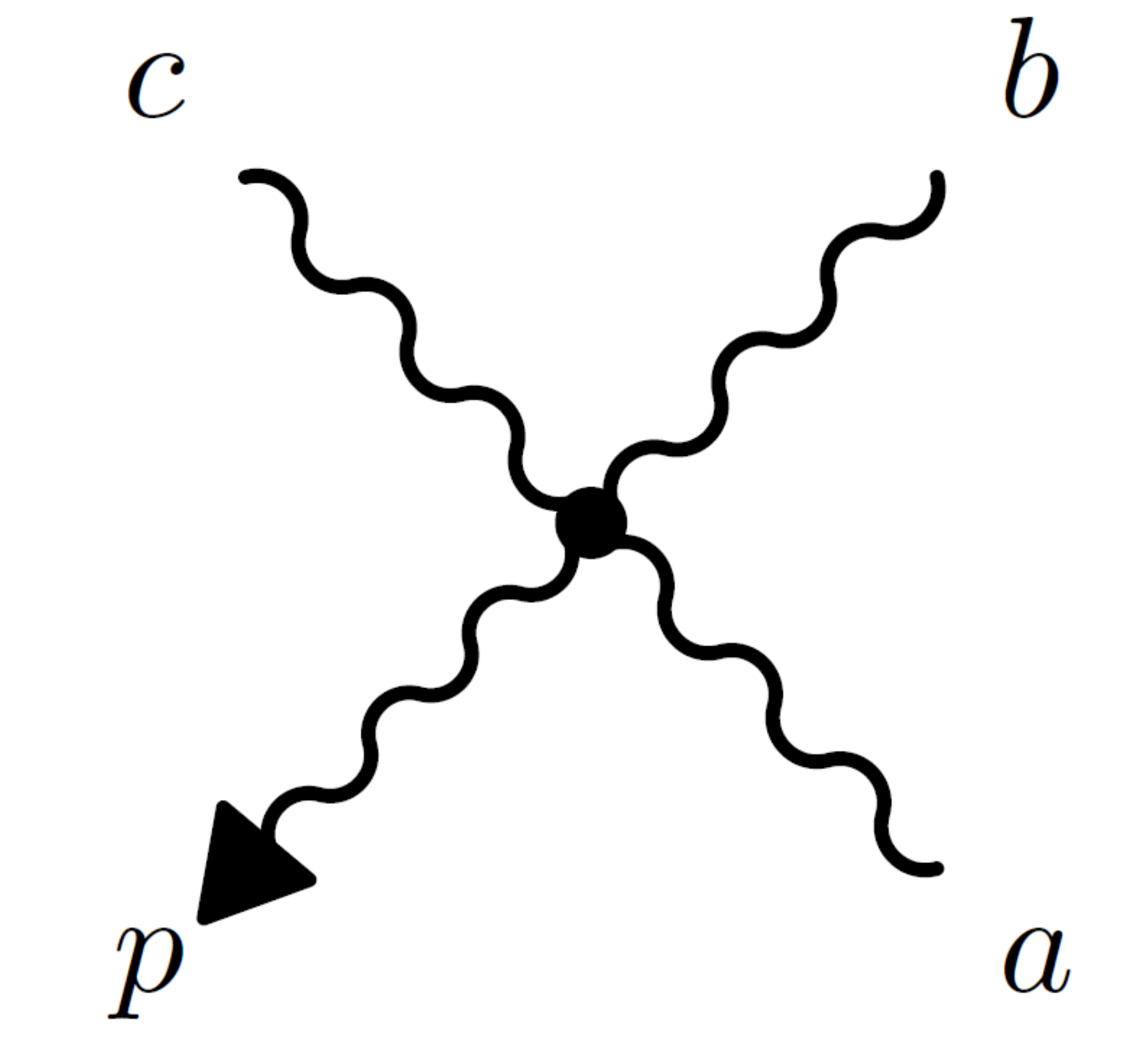}}\right|\raisebox{-0.5cm}{$f^{mpn}\to -i\frac{\kappa}{2\,g}\,(\varepsilon_{p}\cdot k_{m})\, \delta_{mn}$}~.
\ee
for generic indices $m,n$ appearing in the color structure of the four gluon vertex. The final graviton dressing to be considered is the graviton-four-gluon vertex with
an on-shell graviton leg. Here we observe a total decomposition
into inverse propagator dressed legs only
$$
\raisebox{-1.0cm}{\includegraphics[height=2cm]{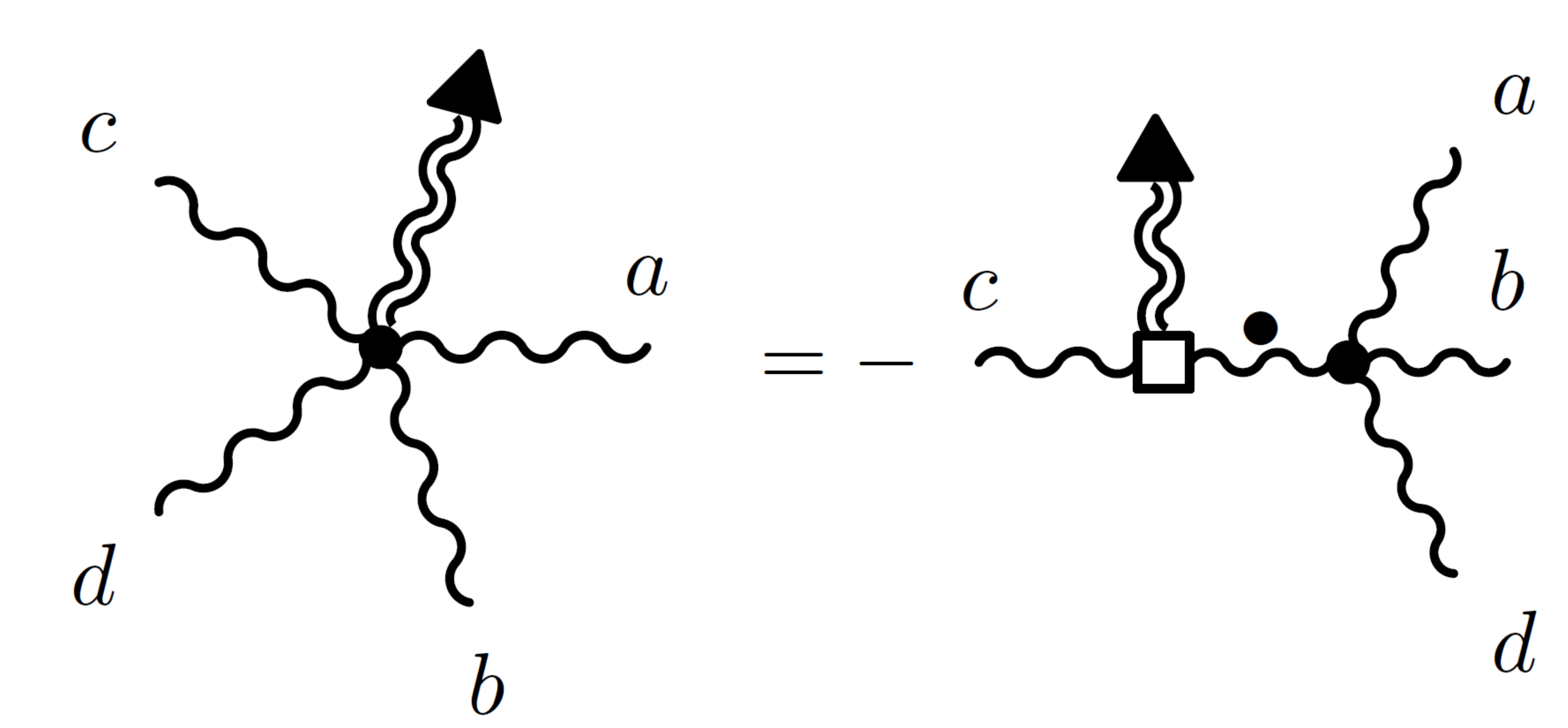}}  \raisebox{-0.1cm}{~~-~\text{($a$, $b$, $d$, $c$) cyclic}\, ,}
$$
and no occurrence of a higher order effective graviton-four-gluon vertex.

Turning to the quark-graviton interactions we observe a similar pattern. The
graviton-quark-anti-quark vertex, the graviton leg being on-shell, is already the effective vertex, now with a $T^{p}$ replacement rule
\be
\label{CCid3}
\left.
\raisebox{-0.87cm}{\includegraphics[height=2cm]{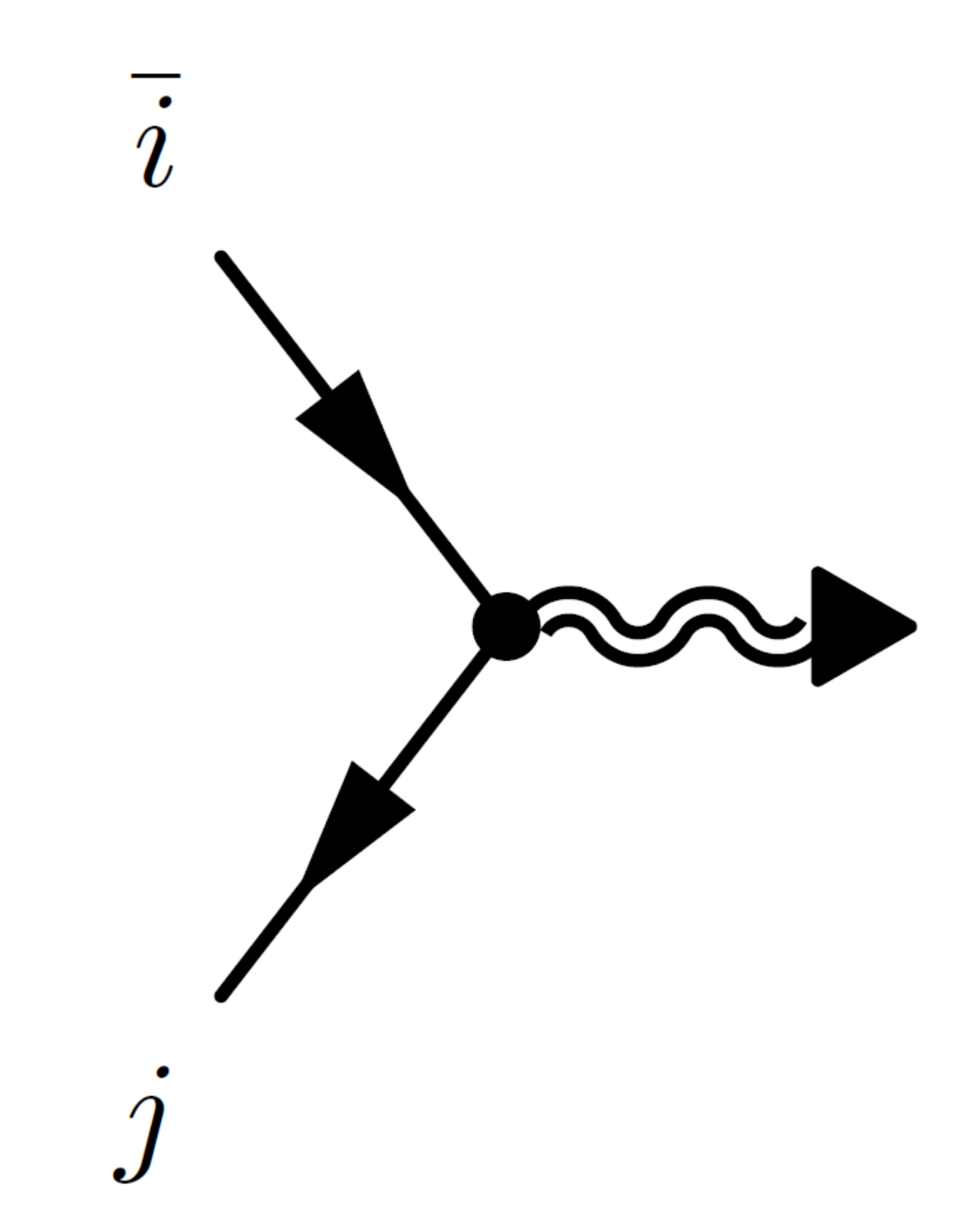}}  = 
\raisebox{-0.87cm}{\includegraphics[height=2cm]{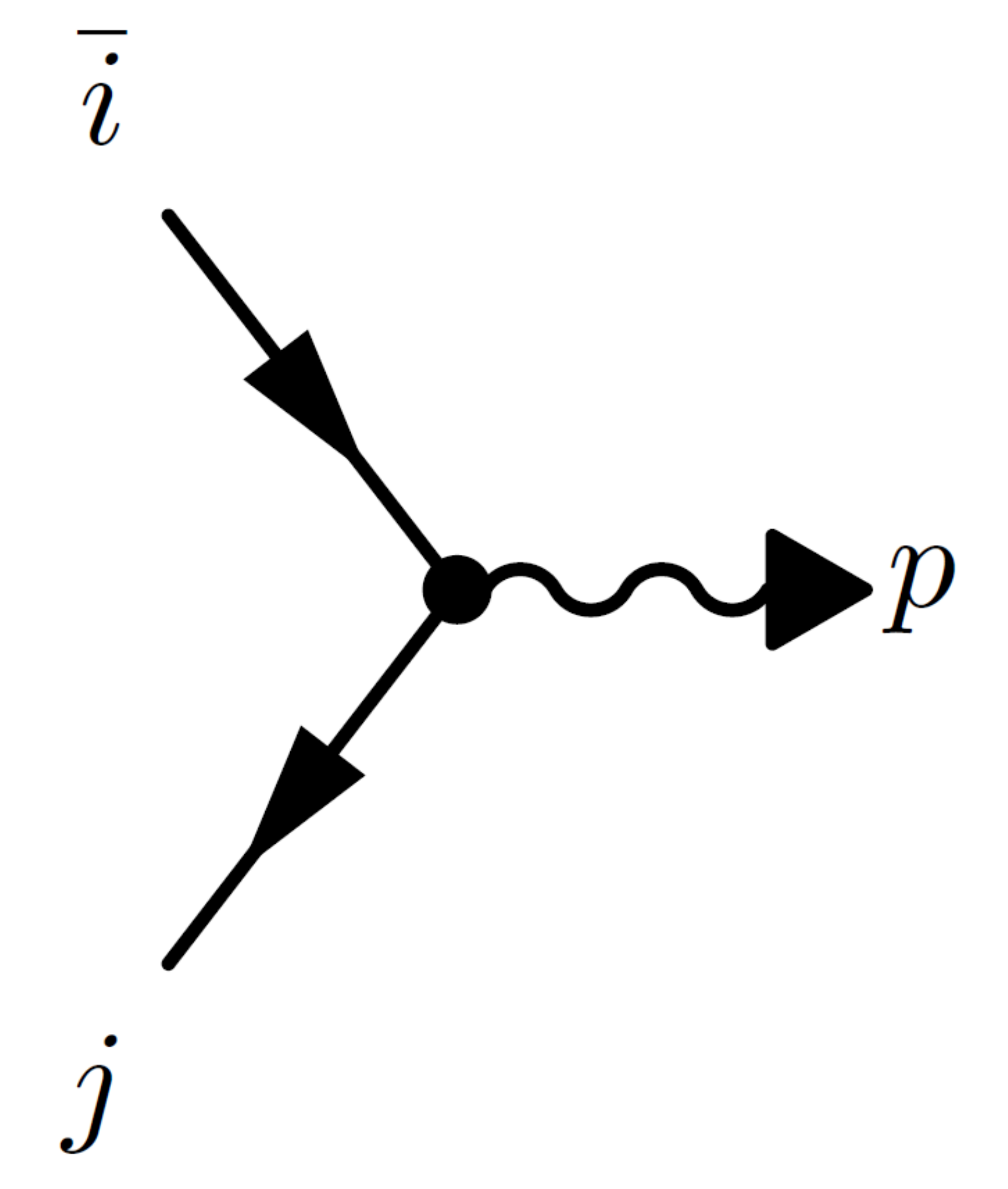}}\right|\raisebox{-0.5cm}{$(T^p)_{j\bar i}\to \frac{\kappa}{2\,g}\,(\varepsilon_{p}\cdot k_{j})\, \delta_{mn}$}~,
\ee
whereas the graviton-gluon-quark-anti-quark vertex gives rise to inverse propagator
legs only %similar to the graviton-four-gluon vertex
$$
\includegraphics[height=1.8cm]{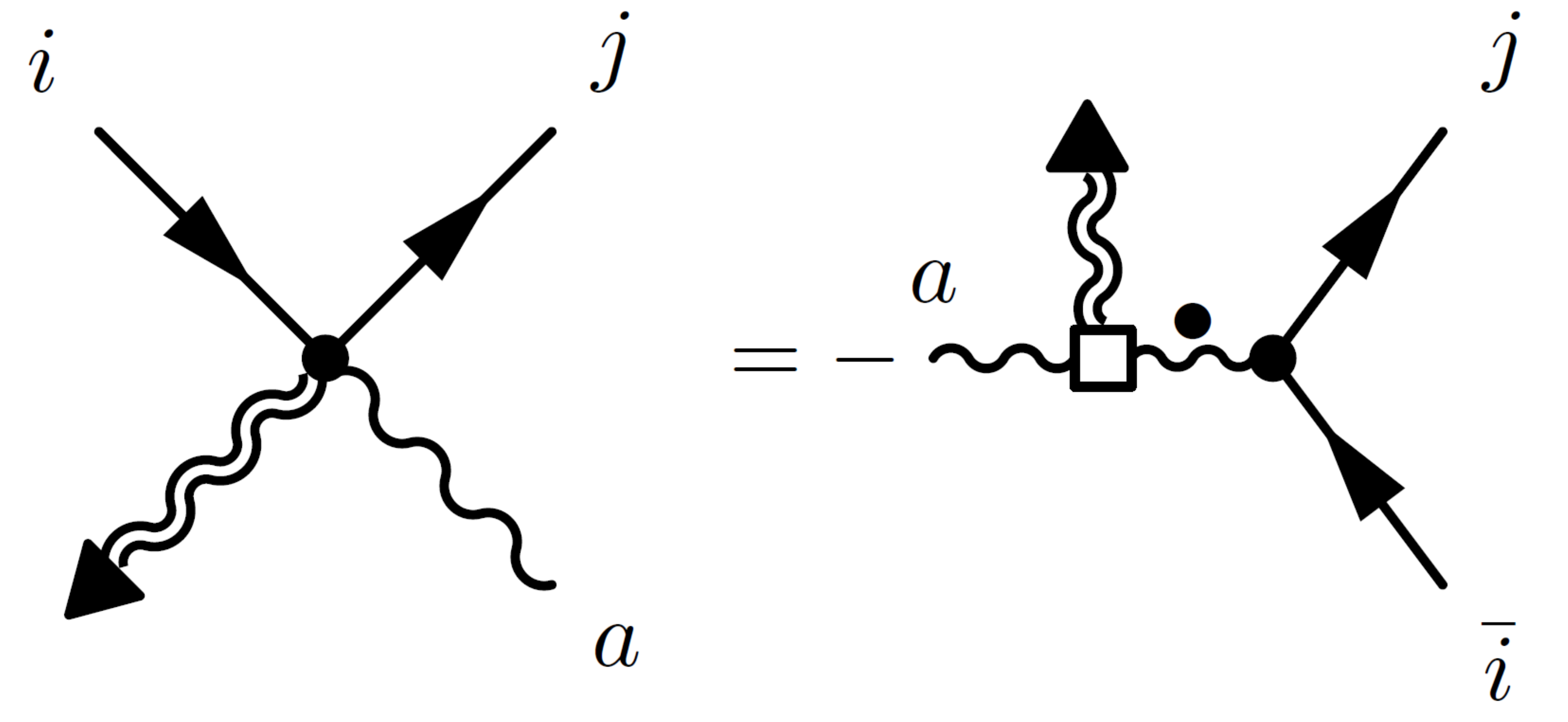}~.
$$
Putting all these insights together we see that in the sum over all possible
attachments of the on-shell graviton leg to the vertices and propagators of the lower point
QCD amplitude, all the inverse propagator (dot on top the gluon leg) terms cancel out and
the single-graviton EQCD amplitude may be evaluated by only considering the 
effective vertices of eqns.~\ref{CCid1}, \eqn{CCid2}, \eqn{CCid3}, which are all generated by the color-kinematic replacement rule 
\be
\mathcal{R}_{p}:= \{(T^{p})_{j\bar{i}}\to \frac{\kappa}{2\,g}(\varepsilon_{p}\cdot k_{j})\, \delta_{j\bar{i}}
\,,\,  \tilde{f}^{apb}%i T^{a_{p}}_{cb}
\to \frac{\kappa}{2g}(\varepsilon_{p}\cdot k_{a})\, \delta_{ab} \}~,
\ee
from the sum over all attachments of a gluon to the same QCD amplitude. This insight entails immediately
the relation for the colored amplitude
\be
\label{Ampreplacement}
\Big.
\cA^{\text{tree}}_{n,k;1}= \cA^{\text{tree}}_{n+1,k,0}
\Big|_{R_{p}} .
\ee

\section{Color-kinematic replacement}

What remains to be understood is how our color-kinematic replacement rule acts on
the MJO-color basis factors $C(\underbar{1},\bar{2},\sigma)$. 
Here we find the key relation
\be
\label{CC-replacement}
C(\underbar{1},\bar{2},\sigma,p,\sigma')
\Bigr |_{\mathcal{R}_{a_{p}}}
= \frac{\kappa}{2g}
(\varepsilon_{p}\cdot k_{2\sigma})\,  C(\underbar{1},\bar{2}, \sigma,\sigma') 
\ee
where $k_{2\sigma} := k_{2}+k_{\sigma_{1}}+ \ldots k_{\sigma_{b}}$ with $b$ being
the length of $\sigma$. To prove this relation we closely follow a proof of a color factor symmetry found in \cite{Brown:2016hck}. In fact, this color factor symmetry may be understood as a direct corollary of our relation \eqn{EQCD-claim} under a gauge transformation  $\varepsilon_{p}^{\mu}\to  p^{\mu}$. We first prove \eqn{CC-replacement} for vanishing 
$\sigma$. Here with \eqn{CC-def} and using the bracket notation we have
\be
C(\underbar{1},\bar{2},p,\sigma')= \{2|\widebar{T}^{a_{p}}\bar{\Omega}|1\}\ldots
=(-)^{\Omega+1}\{1|\Omega \, T^{a_{p}}|2\} \ldots
\ee
where $\Omega$ is determined by the Dyck word of $\sigma$. One also defines  $\widebar{T}^{a}_{\bar{j}i}:= -T^{a}_{i\bar{j}}$  in analogy to $\tilde{f}^{abc}=-\tilde{f}^{cba}$.
With $T_{a_{p}}|2\}|_{\mathcal{R}_{a_{p}}}\to - |2\}\, \frac{\kappa}{2g}(\varepsilon_{p}\cdot k_{2})$  
we indeed find \eqn{CC-replacement} for vanishing $\sigma$. The non-trivial $\sigma$
case is proven by induction.  Let $c=\sigma_{b}$ denote a gluon leg and write 
$$
C(\underbar{1},\bar{2},\sigma_{1}\ldots\sigma_{b},p,\sigma_{b+1}\ldots \sigma_{n-1})=:C_{\ldots|cp|\ldots}\, .
$$
We now assume that \eqn{CC-replacement} is true for $C_{\ldots|pc|\ldots}$. Zooming
in on the position of the insertion we may write
\be
C_{\ldots|cp|\ldots}= \ldots  \Xi^{c}_{l}\, \Xi^{p}_{l}\ldots 
\ee
as the two gluons are next to each other they appear at the same level of nestedness.
Consider now the commutator of $C_{\ldots|cp|\ldots}$ and $C_{\ldots|pc|\ldots}$
\be
C_{\ldots|cp|\ldots}- C_{\ldots|pc|\ldots} =\ldots  [\Xi^{c}_{l},  \Xi^{p}_{l}]\ldots 
= \ldots\tilde{f}^{cpd}\Xi^{d}_{l}\ldots
\ee
Hence applying the color-kinematic replacement rule one has
\be
C_{\ldots|cp|\ldots}- C_{\ldots|pc|\ldots}\Bigr|_{\mathcal{R}_{a_{p}}}=\frac{\kappa}{2g}(\varepsilon_{p}\cdot k_{c})\, C_{\ldots|c|\ldots} \, .
\label{CC-prealgebra}
\ee
By assumption we have $C_{\sigma|pc|\ldots}|_{\mathcal{R}_{a_{p}}}=(\varepsilon_{p}\cdot k_{2\,\sigma})\, C_{\sigma|c|\ldots}$
with $\sigma=\{\sigma_{1}\ldots\sigma_{b-1}\}$ and the induction step is performed.
We note that the relation \eqn{CC-prealgebra} points towards a kinematical algebra
structure
\be
([\Xi^{c}_{l},\Xi^{p}_{l}])_{i\bar{j}}\Bigr|_{\mathcal{R}_{a_{p}}}  =
\frac{\kappa}{2g}(\varepsilon_{p}\cdot k_{c})\, (\Xi^{c}_{l})_{i\bar{j}}\, .
\label{CC-algebra}
\ee
The case where the leg $c=\sigma_{b}$ prior to $p$ is a quark or anti-quark works analogously and is proven in the appendix.
This then completes the proof of the
claimed relation  \eqn{EQCD-claim} relating single graviton EQCD amplitudes to QCD ones.

\begin{figure}
 \includegraphics[width=1.2cm]{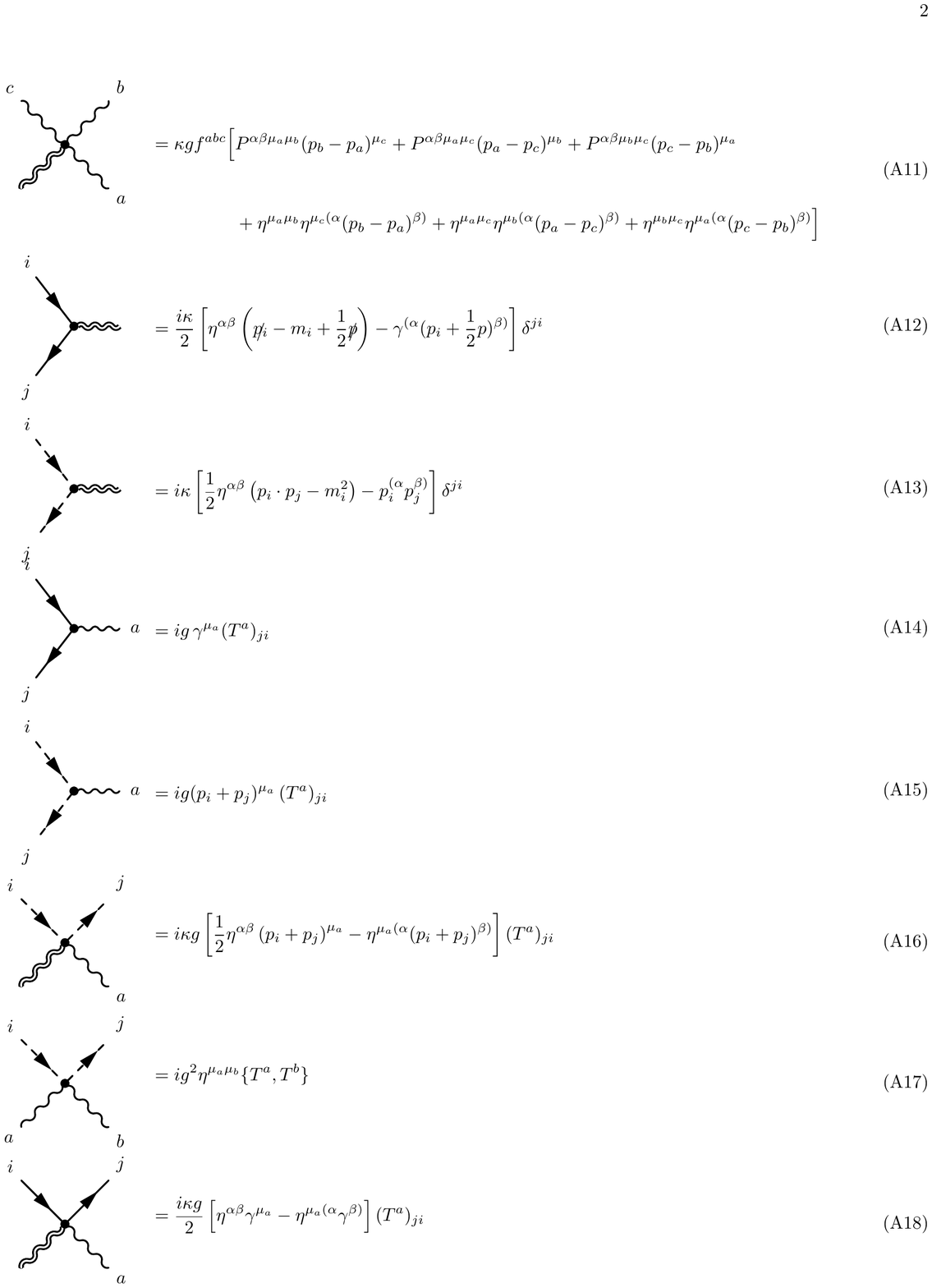}\quad
 \raisebox{0.15cm}{
  \includegraphics[width=1.2cm]{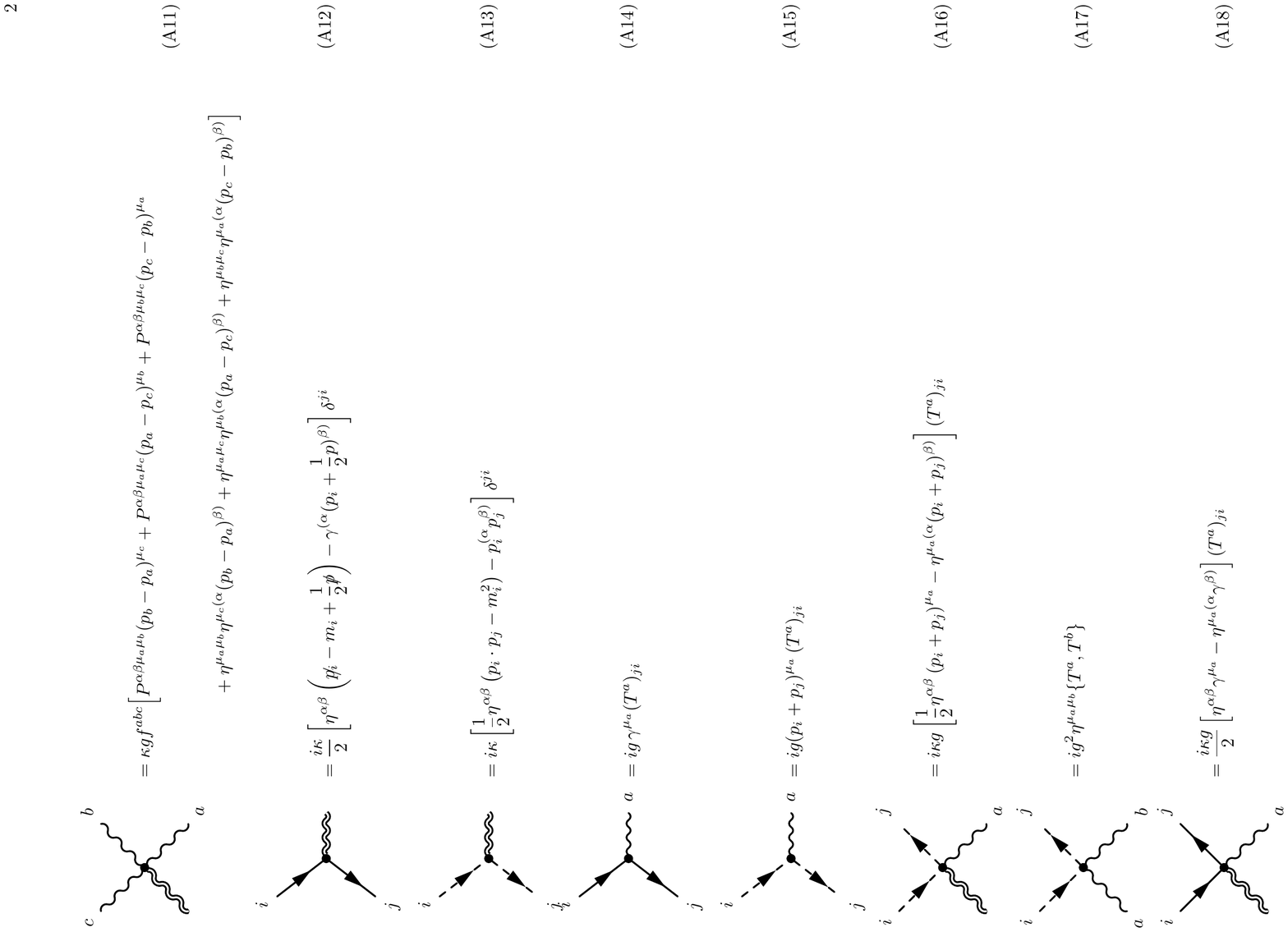}}\quad
  \raisebox{0.1cm}{\includegraphics[width=1.4cm]{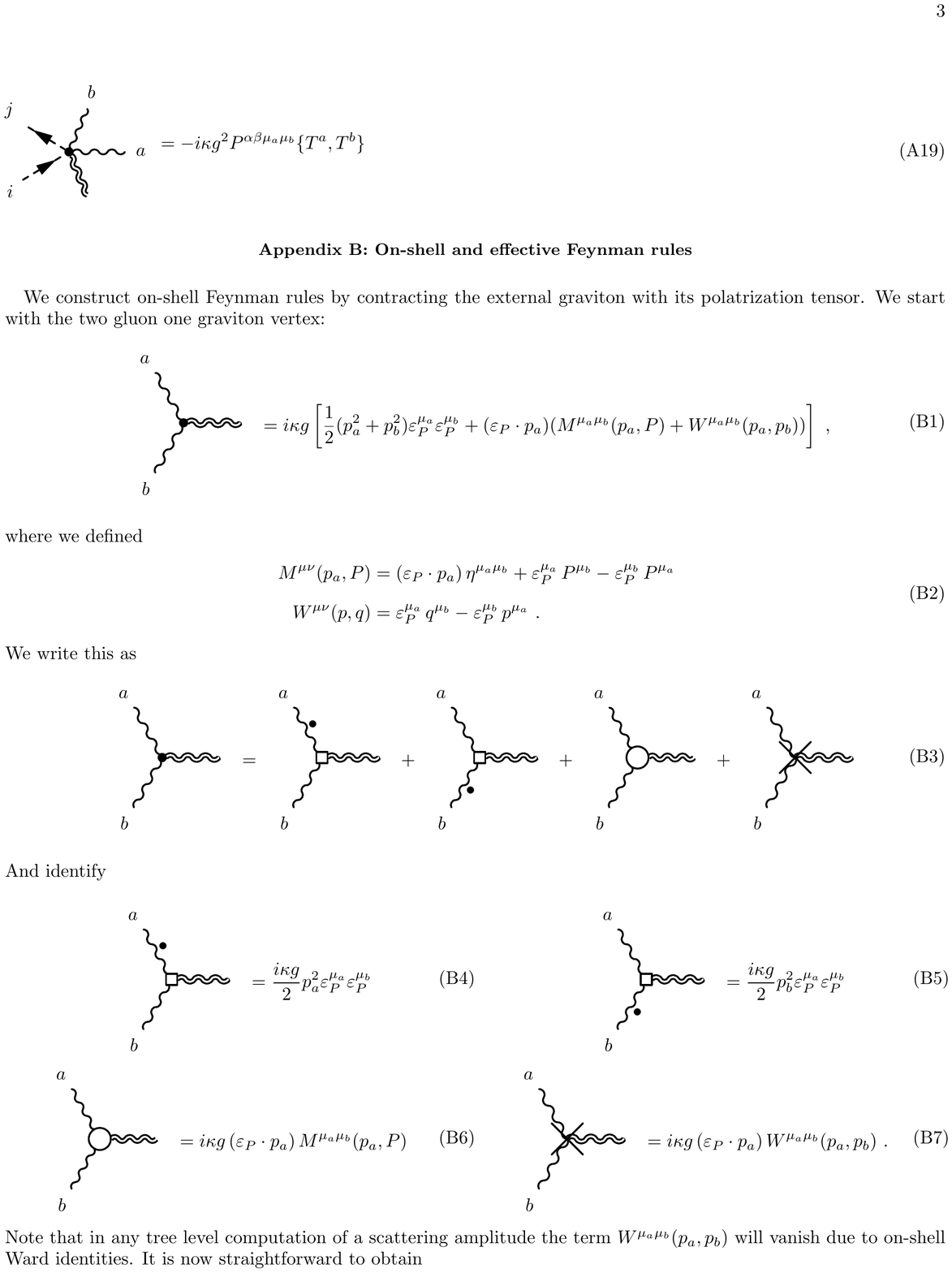}}\quad
 \caption{\label{fig:sEQCD-rules} Single graviton Einstein-sQCD vertices.\\ The dashed lines
 denotes the color charged scalar.}
  \end{figure}

\section{Scalar Matter}

Our central result eqn.~\eqn{EQCD-claim} is in fact universal. It also applies to a theory
of gravitationally minimally interacting massive, color charged scalars, Einstein-Scalar-QCD. Clearly, the color structure of a Scalar-QCD amplitude will be captured by the MJO basis as well. The relevant single
graviton vertices are collected in FIG.~\ref{fig:sEQCD-rules}.
One again shows that the emission of a single graviton from a multi-scalar-gluon
or even multi-scalar-gluon-quark process reduces to coupling the graviton with effective vertices similar to the QCD case (see appendix). I.e.~the gluon `upgrading' relation \eqn{Ampreplacement} holds analogously
and together with the MJO basis reduction \eqn{CC-replacement} this proves \eqn{EQCD-claim}
also for scalar matter.

An interesting question concerns the general matter theories with Yukawa and $\phi^{4}$
(or $\phi^{3}$) couplings (in 4d). Here the issue of a minimal color basis is to be settled. However, considering single graviton emission
processes our diagrammatic argument is straightforwardly extended to these couplings as well:
There are no on-shell graviton couplings to the Yukawa and $\phi^{n}$ vertices. Hence the
gluon upgrade relation \eqn{Ampreplacement} holds here as well.

\section{Discussion}

In a sense the key statement of this letter is that for every BCJ-relation there
is an associated single graviton amplitude: The BCJ-relation is nothing but the gauge invariance of this graviton. What about multiple gravitons?
In \cite{Fu:2017uzt} the single graviton EYM relation was used as a seed for an all graviton
multiplicity reduction of the single trace EYM amplitudes to pure YM ones. The only ingredients were gauge invariance and BCJ relations along with a structural assumption on the form
of the higher graviton amplitudes. If one accepts this assumption also for our general matter
case the multi-graviton, single trace formulae of \cite{Fu:2017uzt} directly generalize. The situation for multi color traces is less clear to us. Finally, we believe that 
 \eqn{CC-algebra} points towards a kinematical algebra structure consisting of a ``mixed'' color-kinematic representation which should be understood in more detail. All
 our results are dimension independent. 
 We also checked our single graviton relation against known amplitudes involving
 massive quarks \cite{Naculich:2014naa} and massive scalars \cite{Nandan:2018ody} in the literature. Moreover, it is possible to extend our techniques to loop level ampitudes with one external on-shell graviton and no internal gravitons (leading order in $\kappa$). It is still possible to use the effective vertex notation except for the case where one inserts a graviton directly into a gluon loop. There it is not possible to use Ward identities and one additional term appears. Still, the new term will be of the same analytic form as the insertion of a gluon to the same gluon loop up to a factor of two. It would be nice to obtain a decomposition similar to \eqref{EQCD-claim} for higher loop level integrands using the techniques provided here.  

\section*{Acknowledgements}
We thank C.~Duhr and A.~Ochirov for discussions. Moreover,
we would like to thank the CERN theory department were this work was completed
and the KITP Santa Barbara where this work was initiated
for hospitality.
This research was supported in part by the National Science Foundation under Grant No. NSF PHY17-48958.

\bibliographystyle{apsrev4-1}
\bibliography{soft}

\newpage

\onecolumngrid
\appendix

\section*{Appendix}
%\section*{Supplemental material}

\subsection{Feynman Rules for Einstein-Yang-Mills-Scalar Theory coupled to Quarks}\label{A}
The theory under consideration is defined by the Lagrangian
\bal{
\mathcal L =\mathcal{L}_{QCD}+ \mathcal{L}_{sQCD}+ \mathcal{L}_{YM} + \mathcal{L}_{EH} + \mathcal{L}_{GF}+\mathcal{L}_{Ghosts}~,
}
with the usual definitions 
\bal{\mathcal L_{YM}=-\frac{1}{4}\,\sqrt{-g}\,g^{\mu\rho}\,g^{\nu\sigma}\,F_{\mu\nu}^a\,F_{\rho\sigma}^a}
\bal{\mathcal L_{QCD} = \sqrt{-g}\,\bar\psi\left(i\slashed{D} - m_{\psi}\right)\psi} 
\bal{\mathcal L_{sQCD} = \sqrt{-g}\,\left(g^{\mu\nu}\,(D_\mu\,\phi)^{\dagger}\,(D_\nu\,\phi) - m_{\phi}\,\phi^\dagger\,\phi\right)}
\bal{\mathcal L_{EH}=\frac{2}{\kappa^2}\,\sqrt{-g}\,g^{\mu\nu}\,R_{\mu\nu}~,}
where $R_{\mu\nu}$ is the usual Ricci tensor, $\kappa$ is the gravitational coupling constant and $D^\mu_{ij} = \partial_\mu\, \delta_{ij} - i\,g\,A^\mu_a\,(T^a)_{ij}$ is the covariant derivative with $g$ being the YM coupling. Furthermore, we take the textbook normalization of the SU(N) generators
\bal{
\Tr(T^a\,T^b) = \frac{1}{2}\,\delta^{ab}~.
} 
We consider the above theory in an expansion around the Minkowski vacuum
\bal{
g_{\mu\nu}= \eta_{\mu\nu} + \kappa\,h_{\mu\nu}~,}
and identify $h_{\mu\nu}$ as the graviton. We choose the Feynman resp.~de Donder gauge fixings
\bal{
\mathcal{L}_{GF} = -\frac{1}{2}\,\left(\partial_\mu\,A^\mu_a\right)^2 + \left(\partial^\nu\,h_{\mu\nu} -\frac{1}{2}\,\partial_\mu\,h^\alpha_\alpha\right)^2~\, .
}
The form of the ghost Lagrangian is not needed since we exclusively work at tree level. The interested reader is referred to \cite{TheoThesis}.\\
The graviton is understood to carry momentum $P$, polarization $\varepsilon^{\alpha\beta}=\varepsilon_P^\alpha\,\varepsilon_P^\beta$ and Lorentz indices $\alpha\beta$. All momenta are ingoing.  Gluons are denoted as wiggly lines, gravitons as double wiggly lines, fermions as solid arrows and scalars as dashed arrows.\\\\
\begin{fmffile}{EINZELBILD}
	\begin{minipage}{0.5\textwidth}
	\bal{\raisebox{-0.9cm}{
		\begin{fmfgraph*}(60,60)
		\fmfleft{a1}
		\fmfright{a2}
		\fmf{photon, label=$P$}{a1,a2}
		\fmfv{label=$a$,l.a=90}{a1}
		\fmfv{label=$b$,l.a=90}{a2}	
	\end{fmfgraph*}}\quad=\frac{-i\,\eta_{\mu_a\mu_b}\,\delta^{ab}}{P^2}}
\end{minipage}
\begin{minipage}{0.5\textwidth}
	\bal{\raisebox{-0.9cm}{
		\begin{fmfgraph*}(60,60)
			\fmfleft{a1}
			\fmfright{a2}
			\fmf{dbl_wiggly, label=$P$}{a1,a2}
			\fmfv{label=$\alpha\beta$,l.a=90}{a1}
			\fmfv{label=$\gamma\delta$,l.a=90}{a2}	
	\end{fmfgraph*}}\quad=\frac{i\,P_{\alpha\beta\gamma\delta}}{P^2}}
\end{minipage}
	\begin{minipage}{0.5\textwidth}
		\bal{\raisebox{-0.9cm}{
				\begin{fmfgraph*}(60,60)
					\fmfleft{a1}
					\fmfright{a2}
					\fmf{fermion, label=$P$}{a1,a2}
					\fmfv{label=$\bar i$,l.a=90}{a1}
					\fmfv{label=$j$,l.a=90}{a2}	
			\end{fmfgraph*}}\quad=\frac{i\,\delta^{j\bar i }}{\slashed{P}-m_{\psi}}}
	\end{minipage}
\begin{minipage}{0.5\textwidth}
	\bal{\raisebox{-0.9cm}{
			\begin{fmfgraph*}(60,60)
				\fmfleft{a1}
				\fmfright{a2}
				\fmf{scalar, label=$P$}{a1,a2}
				\fmfv{label=$\bar i$,l.a=90}{a1}
				\fmfv{label=$j$,l.a=90}{a2}	
		\end{fmfgraph*}}\quad=\frac{i\,\delta^{j\bar i }}{P^2-m_{\phi}^2}}
\end{minipage}\\
where we defined 
\bal{
P_{\alpha\beta\gamma\delta}=\eta_{\alpha(\beta}\eta_{\gamma)\delta} - \frac{1}{d-2}\eta_{\alpha\beta}\eta_{\gamma\delta}\quad,\quad \eta_{\alpha(\gamma}\eta_{\delta)\beta} = \frac{1}{2}\left(\eta_{\alpha\gamma}\,\eta_{\delta\beta}+\eta_{\alpha\delta}\,\eta_{\gamma\beta}\right)
}
	The interaction vertices with at most one graviton read \cite{TheoThesis}\\
	\begin{fleqn}
	\bal{
		\raisebox{-0.75cm}{
			\begin{fmfgraph*}(60,48)
				\fmfpen{thin}
				\fmfsurroundn{h}{3}
				\fmfdot{a1}
				\fmf{photon,tension=0.5}{a1,h1}
				\fmf{photon,tension=0.5}{a1,h2}
				\fmf{photon,tension=0.5}{a1,h3}
				\fmfv{label=$b$}{h2}
				\fmfv{label=$c$}{h3}
				\fmfv{label=$a$}{h1}
		\end{fmfgraph*}}
		\quad~~
		=gf^{abc}\left[\eta^{\mu_a\mu_b}(p_a-p_b)^{\mu_c} + \eta^{\mu_b\mu_c}(p_b-p_c)^{\mu_a} + \eta^{\mu_c\mu_a}(p_c-p_a)^{\mu_b}\right]}
	\\	
	\bal{
		\raisebox{-0.75cm}{
			\begin{fmfgraph*}(60,48)
				\fmfpen{thin}
				\fmfleft{h1,h2}
				\fmfright{h3,h4}
				\fmfdot{a1}
				\fmf{photon,tension=0.5}{a1,h1}
				\fmf{photon,tension=0.5}{a1,h2}
				\fmf{photon,tension=0.5}{a1,h3}
				\fmf{photon,tension=0.5}{a1,h4}
				\fmfv{label=$b$}{h4}
				\fmfv{label=$c$}{h1}
				\fmfv{label=$d$}{h3}
				\fmfv{label=$a$}{h2}
		\end{fmfgraph*}}\quad~~=-ig^2\Big[&f^{abe}f^{cde}\left(\eta^{\mu_a\mu_c}\eta^{\mu_b\mu_d}-\eta^{\mu_a\mu_d}\eta^{\mu_b\mu_c}\right)
		+f^{ace}f^{bde}\left(\eta^{\mu_a\mu_b}\eta^{\mu_c\mu_d}-\eta^{\mu_a\mu_d}\eta^{\mu_b\mu_c}\right)
		\\[-1\baselineskip]&+f^{ade}f^{bce}\left(\eta^{\mu_a\mu_b}\eta^{\mu_c\mu_d}-\eta^{\mu_a\mu_c}\eta^{\mu_b\mu_d}\right)\Big]}
	\bal{
		\raisebox{-0.9cm}{
			\begin{fmfgraph*}(60,60)
				\fmfpen{thin}
				\fmfsurroundn{h}{5}
				\fmfdot{a1}
				\fmf{dbl_wiggly,tension=0.5}{a1,h2}
				\fmf{photon,tension=0.5}{a1,h3}
				\fmf{photon,tension=0.5}{a1,h5}
				\fmf{photon,tension=0.5}{a1,h1}
				\fmf{photon,tension=0.5}{a1,h4}
				\fmfv{label=$b$}{h5}
				\fmfv{label=$c$}{h3}
				\fmfv{label=$d$}{h4}
				\fmfv{label=$a$}{h1}
		\end{fmfgraph*}}\quad~~=	(ig^2\kappa&)\Big[f^{abe}f^{cde}(P^{\alpha\beta\mu_a\mu_c}\eta^{\mu_b\mu_d}+P^{\alpha\beta\mu_b\mu_d}\eta^{\mu_a\mu_c}-P^{\alpha\beta\mu_b\mu_c}\eta^{\mu_a\mu_d}-P^{\alpha\beta\mu_a\mu_d}\eta^{\mu_b\mu_c})\\[-\baselineskip]
	&+f^{ace}f^{bde}(P^{\alpha\beta\mu_a\mu_b}\eta^{\mu_c\mu_d}+P^{\alpha\beta\mu_c\mu_d}\eta^{\mu_a\mu_b}-P^{\alpha\beta\mu_c\mu_b}\eta^{\mu_a\mu_d}-P^{\alpha\beta\mu_a\mu_d}\eta^{\mu_b\mu_c})\\
&+f^{ade}f^{bce}(P^{\alpha\beta\mu_a\mu_c}\eta^{\mu_b\mu_d}+P^{\alpha\beta\mu_d\mu_b}\eta^{\mu_a\mu_c}-P^{\alpha\beta\mu_d\mu_c}\eta^{\mu_a\mu_b}-P^{\alpha\beta\mu_a\mu_b}\eta^{\mu_d\mu_c})\Big]}
	\bal{
	\raisebox{-0.9cm}{
		\begin{fmfgraph*}(60,60)
			\fmfpen{thin}
			\fmfsurroundn{h}{3}
			\fmfdot{a1}
			\fmf{dbl_wiggly,tension=0.5}{a1,h1}
			\fmf{photon,tension=0.5}{a1,h2}
			\fmf{photon,tension=0.5}{a1,h3}
			\fmfv{label=$a$}{h2}
			\fmfv{label=$b$}{h3}
	\end{fmfgraph*}}
	\quad~~
	=-i\kappa g \left[P^{\alpha\beta\mu_a\mu_b}\,p_a\cdot p_b + \eta^{\mu_a\mu_b}p_b^{(\alpha}p_a^{\beta)} - \eta^{\mu_a(\alpha}p_b^{\beta)}p_a^{\mu_b} - \eta^{\mu_b(\alpha}p_a^{\beta)}p_b^{\mu_a} + \frac{1}{2}\eta^{\alpha\beta}p_a^{\mu_a} p_b^{\mu_b}\right]}\\
	\bal{
	\raisebox{-0.75cm}{
		\begin{fmfgraph*}(60,48)
			\fmfpen{thin}
			\fmfleft{h1,h2}
			\fmfright{h3,h4}
			\fmfdot{a1}
			\fmf{dbl_wiggly,tension=0.5}{a1,h1}
			\fmf{photon,tension=0.5}{a1,h2}
			\fmf{photon,tension=0.5}{a1,h3}
			\fmf{photon,tension=0.5}{a1,h4}
			\fmfv{label=$b$}{h4}
			\fmfv{label=$c$}{h2}
			\fmfv{label=$a$}{h3}
	\end{fmfgraph*}}\quad~~=\kappa g f^{abc}\Big[&P^{\alpha\beta\mu_a\mu_b}(p_b-p_a)^{\mu_c}+P^{\alpha\beta\mu_a\mu_c}(p_a-p_c)^{\mu_b}+P^{\alpha\beta\mu_b\mu_c}(p_c-p_b)^{\mu_a} \\[-\baselineskip]&+ \eta^{\mu_a\mu_b}\eta^{\mu_c(\alpha}(p_b-p_a)^{\beta)} + \eta^{\mu_a\mu_c}\eta^{\mu_b(\alpha}(p_a-p_c)^{\beta)}+ \eta^{\mu_b\mu_c}\eta^{\mu_a(\alpha}(p_c-p_b)^{\beta)}\Big]}
	\bal{
	\raisebox{-0.9cm}{
		\begin{fmfgraph*}(60,60)
			\fmfpen{thin}
			\fmfsurroundn{h}{3}
			\fmfdot{a1}
			\fmf{dbl_wiggly,tension=0.5}{a1,h1}
			\fmf{quark,tension=0.3}{h2,a1}
			\fmf{quark,tension=0.3}{a1,h3}
			\fmfv{label=$\bar i$}{h2}
			\fmfv{label=$j$}{h3}
	\end{fmfgraph*}}
	\quad~~
	=\frac{i\kappa}{2}\left[\eta^{\alpha\beta}\left(\slashed{p}_{i} - m_{\psi} + \frac{1}{2}\slashed{p}\right) - \gamma^{(\alpha}(p_{ i}+\frac{1}{2}p)^{\beta)}\right]\delta^{j \bar i}}
\\
	\bal{
	\raisebox{-0.9cm}{
		\begin{fmfgraph*}(60,60)
			\fmfpen{thin}
			\fmfsurroundn{h}{3}
			\fmfdot{a1}
			\fmf{dbl_wiggly,tension=0.5}{a1,h1}
			\fmf{scalar,tension=0.3}{h2,a1}
			\fmf{scalar,tension=0.3}{a1,h3}
			\fmfv{label=$\bar i$}{h2}
			\fmfv{label=$j$}{h3}
	\end{fmfgraph*}}
	\quad~~
	=i\kappa\left[\frac{1}{2}\eta^{\alpha\beta}\left(-p_i\cdot p_j - m_i^2\right) + p_i^{(\alpha}p_j^{\beta)}\right]\delta^{j\bar i}}\\
	\bal{
		\raisebox{-0.9cm}{
			\begin{fmfgraph*}(60,60)
				\fmfpen{thin}
				\fmfsurroundn{h}{3}
				\fmfdot{a1}
				\fmf{photon,tension=0.5}{a1,h1}
				\fmf{quark,tension=0.3}{h2,a1}
				\fmf{quark,tension=0.3}{a1,h3}
				\fmfv{label=$\bar i$}{h2}
				\fmfv{label=$j$}{h3}
				\fmfv{label=$a$}{h1}
		\end{fmfgraph*}}
		\quad~~
		=ig\,\gamma^{\mu_a}(T^a)_{j\bar i}}\\
	\bal{
	\raisebox{-0.9cm}{
		\begin{fmfgraph*}(60,60)
			\fmfpen{thin}
			\fmfsurroundn{h}{3}
			\fmfdot{a1}
			\fmf{photon,tension=0.5}{a1,h1}
			\fmf{scalar,tension=0.3}{h2,a1}
			\fmf{scalar,tension=0.3}{a1,h3}
			\fmfv{label=$\bar i$}{h2}
			\fmfv{label=$j$}{h3}
			\fmfv{label=$a$}{h1}
	\end{fmfgraph*}}
	\quad~~
	=ig(p_i - p_j)^{\mu_a}\,(T^a)_{j\bar i}}
\\
	\bal{
	\raisebox{-0.9cm}{
		\begin{fmfgraph*}(60,48)
			\fmfpen{thin}
			\fmfleft{h1,h2}
			\fmfright{h3,h4}
			\fmfdot{a1}
			\fmf{photon,tension=0.5}{h3,a1}
			\fmf{dbl_wiggly,tension=0.5}{a1,h1}
			\fmf{scalar,tension=0.5}{h2,a1}
			\fmf{scalar,tension=0.5}{a1,h4}
			\fmfv{label=$\bar i$}{h2}
			\fmfv{label=$j$}{h4}
			\fmfv{label=$a$}{h3}
	\end{fmfgraph*}}
	\quad~~
	=i\kappa g \left[\frac{1}{2}\eta^{\alpha\beta}\left(p_i - p_j\right)^{\mu_a} - \eta^{\mu_a(\alpha}(p_i-p_j)^{\beta)}\right](T^a)_{j\bar i}}
\\
	\bal{
	\raisebox{-0.9cm}{
		\begin{fmfgraph*}(60,48)
			\fmfpen{thin}
			\fmfleft{h1,h2}
			\fmfright{h3,h4}
			\fmfdot{a1}
			\fmf{photon,tension=0.5}{h3,a1}
			\fmf{photon,tension=0.5}{a1,h1}
			\fmf{scalar,tension=0.5}{h2,a1}
			\fmf{scalar,tension=0.5}{a1,h4}
			\fmfv{label=$\bar i$}{h2}
			\fmfv{label=$j$}{h4}
			\fmfv{label=$a$}{h1}
			\fmfv{label=$b$}{h3}
	\end{fmfgraph*}}
\quad~~
	=ig^2\eta^{\mu_a\mu_b}\{T^a,T^b\}}
\\
	\bal{
	\raisebox{-0.9cm}{
		\begin{fmfgraph*}(60,48)
			\fmfpen{thin}
			\fmfleft{h1,h2}
			\fmfright{h3,h4}
			\fmfdot{a1}
			\fmf{photon,tension=0.5}{h3,a1}
			\fmf{dbl_wiggly,tension=0.5}{a1,h1}
			\fmf{fermion,tension=0.5}{h2,a1}
			\fmf{fermion,tension=0.5}{a1,h4}
			\fmfv{label=$\bar i$}{h2}
			\fmfv{label=$j$}{h4}
			\fmfv{label=$a$}{h3}
	\end{fmfgraph*}}
\quad~~
	=\frac{i\kappa g}{2} \left[\eta^{\alpha\beta}\gamma^{\mu_a} - \eta^{\mu_a(\alpha}\gamma^{\beta)}\right](T^a)_{j\bar i}}
\\
	\bal{
	\raisebox{-0.9cm}{
		\begin{fmfgraph*}(60,48)
			\fmfpen{thin}
			\fmfsurround{h1,h2,h3,h4,h5}
			\fmfdot{a1}
			\fmf{photon,tension=0.5}{h2,a1}
			\fmf{photon,tension=0.5}{a1,h1}
			\fmf{scalar,tension=0.5}{a1,h3}
			\fmf{scalar,tension=0.5}{h4,a1}
			\fmf{dbl_wiggly,tension=0.5}{a1,h5}
			\fmfv{label=$\bar i$}{h4}
			\fmfv{label=$j$}{h3}
			\fmfv{label=$a$}{h1}
			\fmfv{label=$b$}{h2}
	\end{fmfgraph*}}
	\quad~~
	=-i\kappa g^2 P^{\alpha\beta\mu_a\mu_b}\{T^a,T^b\}}
\end{fleqn}
\section{On-shell and effective Feynman rules}
We construct on-shell Feynman rules by contracting one external graviton or gluon with its polarization tensor and using the conditions $\varepsilon_P\cdot P= \varepsilon_P^2 =0$. Graphically we denote the on-shell leg with a triangle. We start with the two gluon one graviton vertex:\\
\bal{
	\raisebox{-0.9cm}{
		\begin{fmfgraph*}(60,60)
			\fmfpen{thin}
			\fmfsurroundn{h}{3}
			\fmfdot{a1}
			\fmf{dbl_wiggly,tension=0.5}{a1,h1}
			\fmf{photon,tension=0.5}{a1,h2}
			\fmf{photon,tension=0.5}{a1,h3}
			\fmfv{label=$a$}{h2}
			\fmfv{label=$b$}{h3}
			\fmfv{decor.shape=triangle, decor.filled=full, decor.size=0.15w,d.a=-90}{h1}
	\end{fmfgraph*}}
	\quad
	=i\kappa g\, \delta^{ab}\,\left[\frac{1}{2}(p_a^2 + p_b^2)\varepsilon_P^{\mu_a}\varepsilon_P^{\mu_b} + (\varepsilon_P\cdot p_a)( M^{\mu_a\mu_b} + W^{\mu_a\mu_b})\right]~,}\\\\
where we defined
\bal{
 M^{\mu\nu} &= (\varepsilon_P\cdot p_a)\,\eta^{\mu_a\mu_b} + \varepsilon_P^{\mu_a}\,P^{\mu_b} - \varepsilon_P^{\mu_b}\,P^{\mu_a}\\
  W^{\mu\nu} & = \varepsilon_P^{\mu_a}\,p_{b}^{\mu_b} - \varepsilon_P^{\mu_b}\,
  p_{a}^{\mu_a}~.
}
Moreover we used 
\bal{
p_a\cdot p_b = \frac{1}{2}\left((p_a + p_b)^2 - p_a^2 - p_b^2\right)~,  
}
where the first term in the bracket is $p^2=0$.
We write this as\\
\bal{
	\raisebox{-0.9cm}{
		\begin{fmfgraph*}(60,60)
			\fmfpen{thin}
			\fmfsurroundn{h}{3}
			\fmfdot{a1}
			\fmf{dbl_wiggly,tension=0.5}{a1,h1}
			\fmf{photon,tension=0.5}{a1,h2}
			\fmf{photon,tension=0.5}{a1,h3}
			\fmfv{label=$a$}{h2}
			\fmfv{label=$b$}{h3}
						\fmfv{decor.shape=triangle, decor.filled=full, decor.size=0.15w,d.a=-90}{h1}
	\end{fmfgraph*}}\quad=
\raisebox{-0.9cm}{
	\begin{fmfgraph*}(60,60)
		\fmfpen{thin}
		\fmfsurroundn{h}{3}
		\fmfv{decor.shape=square, decor.filled=empty,decor.size= 0.1w}{a1}
		\fmf{dbl_wiggly,tension=0.5}{a1,h1}
		\fmf{photon,label=$\bullet$,l.d=0.05cm,tension=0.5}{a1,h2}
		\fmf{photon,tension=0.5}{a1,h3}
		\fmfv{label=$a$}{h2}
		\fmfv{label=$b$}{h3}
					\fmfv{decor.shape=triangle, decor.filled=full, decor.size=0.15w,d.a=-90}{h1}
\end{fmfgraph*}}
 \quad + 
\raisebox{-0.9cm}{
	\begin{fmfgraph*}(60,60)
		\fmfpen{thin}
		\fmfsurroundn{h}{3}
		\fmfv{decor.shape=square, decor.filled=empty,decor.size= 0.1w}{a1}
		\fmf{dbl_wiggly,tension=0.5}{a1,h1}
		\fmf{photon,tension=0.5}{a1,h2}
		\fmf{photon,label=$\bullet$,l.d=0.05cm,tension=0.5}{h3,a1}
					\fmfv{label=$a$}{h2}
								\fmfv{decor.shape=triangle, decor.filled=full, decor.size=0.15w,d.a=-90}{h1}
		\fmfv{label=$b$}{h3}
\end{fmfgraph*}}\quad+ 
\raisebox{-0.9cm}{
	\begin{fmfgraph*}(60,60)
		\fmfpen{thin}
		\fmfsurroundn{h}{3}
		\fmfv{decor.shape=circle, decor.filled=empty,decor.size= 0.2w}{a1}
		\fmf{dbl_wiggly,tension=0.5}{a1,h1}
		\fmf{photon,tension=0.5}{a1,h2}
		\fmf{photon,tension=0.5}{a1,h3}
					\fmfv{decor.shape=triangle, decor.filled=full, decor.size=0.15w,d.a=-90}{h1}
					\fmfv{label=$a$}{h2}
		\fmfv{label=$b$}{h3}
\end{fmfgraph*}}
\quad+
\raisebox{-0.9cm}{
	\begin{fmfgraph*}(60,60)
		\fmfpen{thin}
		\fmfsurroundn{h}{3}
		\fmfv{decor.shape=cross,decor.size= 0.4w}{a1}
		\fmf{dbl_wiggly,tension=0.5}{a1,h1}
		\fmf{photon,tension=0.5}{a1,h2}
		\fmf{photon,tension=0.5}{a1,h3}
					\fmfv{label=$a$}{h2}
		\fmfv{label=$b$}{h3}
					\fmfv{decor.shape=triangle, decor.filled=full, decor.size=0.15w,d.a=-90}{h1}
\end{fmfgraph*}}}\\\\
And identify\\\\\\\\
\begin{minipage}{0.5\textwidth}
	\bal{
		\raisebox{-0.9cm}{
				\begin{fmfgraph*}(60,60)
				\fmfpen{thin}
				\fmfsurroundn{h}{3}
				\fmfv{decor.shape=square, decor.filled=empty,decor.size= 0.1w}{a1}
				\fmf{dbl_wiggly,tension=0.5}{a1,h1}
				\fmf{photon,label=$\bullet$,l.d=0.05cm,tension=0.5}{a1,h2}
				\fmf{photon,tension=0.5}{a1,h3}
							\fmfv{label=$a$}{h2}
				\fmfv{label=$b$}{h3}
					\fmfv{decor.shape=triangle, decor.filled=full, decor.size=0.15w,d.a=-90}{h1}
		\end{fmfgraph*}}\quad= \frac{i\kappa g}{2} p_a^2 \varepsilon_P^{\mu_a}\varepsilon_P^{\mu_b}\,\delta^{ab}}
\end{minipage}
\begin{minipage}{0.5\textwidth}
	\bal{
		\raisebox{-0.9cm}{
				\begin{fmfgraph*}(60,60)
				\fmfpen{thin}
				\fmfsurroundn{h}{3}
				\fmfv{decor.shape=square, decor.filled=empty,decor.size= 0.1w}{a1}
				\fmf{dbl_wiggly,tension=0.5}{a1,h1}
				\fmf{photon,tension=0.5}{a1,h2}
				\fmf{photon,label=$\bullet$,l.d=0.05cm,tension=0.5}{h3,a1}
							\fmfv{label=$a$}{h2}
				\fmfv{label=$b$}{h3}
					\fmfv{decor.shape=triangle, decor.filled=full, decor.size=0.15w,d.a=-90}{h1}
		\end{fmfgraph*}}\quad= \frac{i\kappa g}{2} p_b^2 \varepsilon_P^{\mu_a}\varepsilon_P^{\mu_b}\,\delta^{ab}}
\end{minipage}\\\\\\\\\\
\begin{minipage}{0.5\textwidth}
	\bal{
		\raisebox{-0.9cm}{
				\begin{fmfgraph*}(60,60)
				\fmfpen{thin}
				\fmfsurroundn{h}{3}
				\fmfv{decor.shape=circle, decor.filled=empty,decor.size= 0.2w}{a1}
				\fmf{dbl_wiggly,tension=0.5}{a1,h1}
				\fmf{photon,tension=0.5}{a1,h2}
				\fmf{photon,tension=0.5}{a1,h3}
							\fmfv{label=$a$}{h2}
				\fmfv{label=$b$}{h3}
					\fmfv{decor.shape=triangle, decor.filled=full, decor.size=0.15w,d.a=-90}{h1}
		\end{fmfgraph*}}\quad= i\kappa g\, (\varepsilon_P\cdot p_a)\,M^{\mu_a\mu_b}\,\delta^{ab}\label{eq:effectiveVertexGGGr}}
\end{minipage}
\begin{minipage}{0.5\textwidth}
	\bal{
		\raisebox{-0.9cm}{
			\begin{fmfgraph*}(60,60)
				\fmfpen{thin}
				\fmfsurroundn{h}{3}
				\fmfv{decor.shape=cross,decor.size= 0.4w}{a1}
				\fmf{dbl_wiggly,tension=0.5}{a1,h1}
				\fmf{photon,tension=0.5}{a1,h2}
				\fmf{photon,tension=0.5}{a1,h3}
							\fmfv{label=$a$}{h2}
				\fmfv{label=$b$}{h3}	\fmfv{decor.shape=triangle, decor.filled=full, decor.size=0.15w,d.a=-90}{h1}
		\end{fmfgraph*}}\quad= i\kappa g\, (\varepsilon_P\cdot p_a)\,W^{\mu_a\mu_b}\,\delta^{ab}~.}
\end{minipage}
\vspace{5mm}
\\
Note that in any tree level computation of a scattering amplitude the term $W^{\mu_a\mu_b}(p_a,p_b)$ will vanish due to on-shell Ward identities. It is now straightforward to obtain\\\\
	\bal{
	\raisebox{-0.75cm}{
		\begin{fmfgraph*}(60,48)
			\fmfpen{thin}
			\fmfleft{h1,h2}
			\fmfright{h3,h4}
			\fmfdot{a1}
			\fmf{dbl_wiggly,tension=0.5}{a1,h1}
			\fmf{photon,tension=0.5}{a1,h2}
			\fmf{photon,tension=0.5}{a1,h3}
			\fmf{photon,tension=0.5}{a1,h4}
			\fmfv{label=$b$}{h4}
			\fmfv{label=$c$}{h2}
			\fmfv{label=$a$}{h3}	\fmfv{decor.shape=triangle, decor.filled=full, decor.size=0.15w,d.a=135}{h1}
	\end{fmfgraph*}}\qquad~~=	\raisebox{-0.75cm}{
\begin{fmfgraph*}(60,48)
\fmfpen{thin}
\fmfleft{h1,h2}
\fmfright{h3,h4}
\fmfv{decor.shape=circle, decor.filled=empty,decor.size= 0.2w}{a1}
\fmf{dbl_wiggly,tension=0.5}{a1,h1}
\fmf{photon,tension=0.5}{a1,h2}
\fmf{photon,tension=0.5}{a1,h3}
\fmf{photon,tension=0.5}{a1,h4}
\fmfv{label=$b$}{h4}
\fmfv{label=$c$}{h2}
\fmfv{label=$a$}{h3}	\fmfv{decor.shape=triangle, decor.filled=full, decor.size=0.15w,d.a=135}{h1}
\end{fmfgraph*}}~~-\quad
\raisebox{-0.75cm}{	
	\begin{fmfgraph*}(60,48)
		\fmfpen{thin}
		\fmfleft{i1}
		\fmfright{i2,i3}
		\fmfdot{a1}
		\fmf{photon,tension=1,label=$\bullet$,l.d=0.05cm}{a1,v1}
		\fmf{photon,tension=1}{v1,i1}
		\fmf{photon,tension=1}{a1,v2}
		\fmf{photon,tension=1}{v2,i2}
		\fmf{photon,tension=1}{a1,v3}
		\fmf{photon,tension=1}{v3,i3}
		\fmfv{label=$c$,l.a=90}{i1}
		\fmfv{label=$b$}{i3}
		\fmfv{label=$a$}{i2}
		\fmfsurroundn{h}{361}
		\fmffreeze
		\fmfv{decor.shape=square,decor.filled=empty,decor.size=0.1w}{v1}
		\fmf{dbl_wiggly}{h117,v1}
			\fmfv{decor.shape=triangle, decor.filled=full, decor.size=0.15w}{h117}
\end{fmfgraph*}}
~~-\quad
\raisebox{-0.75cm}{	
	\begin{fmfgraph*}(60,48)
		\fmfpen{thin}
		\fmfleft{i1}
		\fmfright{i2,i3}
		\fmfdot{a1}
		\fmf{photon,tension=1,label=$\bullet$,l.d=0.05cm}{a1,v1}
		\fmf{photon,tension=1}{v1,i1}
		\fmf{photon,tension=1}{a1,v2}
		\fmf{photon,tension=1}{v2,i2}
		\fmf{photon,tension=1}{a1,v3}
		\fmf{photon,tension=1}{v3,i3}
		\fmfv{label=$b$,l.a=90}{i1}
		\fmfv{label=$a$}{i3}
		\fmfv{label=$c$}{i2}
		\fmfsurroundn{h}{361}
		\fmffreeze
		\fmfv{decor.shape=square,decor.filled=empty,decor.size=0.1w}{v1}
		\fmf{dbl_wiggly}{h117,v1}
			\fmfv{decor.shape=triangle, decor.filled=full, decor.size=0.15w}{h117}
\end{fmfgraph*}}
-\quad
%graviton inserted top right gluon leg in ggg vertex
\raisebox{-0.75cm}{	
	\begin{fmfgraph*}(60,48)
		\fmfpen{thin}
		\fmfleft{i1}
		\fmfright{i2,i3}
		\fmfdot{a1}
		\fmf{photon,tension=1,label=$\bullet$,l.d=0.05cm}{a1,v1}
		\fmf{photon,tension=1}{v1,i1}
		\fmf{photon,tension=1}{a1,v2}
		\fmf{photon,tension=1}{v2,i2}
		\fmf{photon,tension=1}{a1,v3}
		\fmf{photon,tension=1}{v3,i3}
		\fmfv{label=$a$,l.a=90}{i1}
		\fmfv{label=$c$}{i3}
		\fmfv{label=$b$}{i2}
		\fmfsurroundn{h}{361}
		\fmffreeze
		\fmfv{decor.shape=square,decor.filled=empty,decor.size=0.1w}{v1}
		\fmf{dbl_wiggly}{h117,v1}
			\fmfv{decor.shape=triangle, decor.filled=full, decor.size=0.15w}{h117}
\end{fmfgraph*}}}\\\\
\bal{
\raisebox{-0.9cm}{
	\begin{fmfgraph*}(60,60)
		\fmfpen{thin}
		\fmfsurroundn{h}{5}
		\fmfdot{a1}
		\fmf{dbl_wiggly,tension=0.5}{a1,h2}
		\fmf{photon,tension=0.5}{a1,h3}
		\fmf{photon,tension=0.5}{a1,h5}
		\fmf{photon,tension=0.5}{a1,h1}
		\fmf{photon,tension=0.5}{a1,h4}
		\fmfv{label=$b$}{h5}
		\fmfv{label=$c$}{h3}
		\fmfv{label=$d$}{h4}
		\fmfv{label=$a$,l.a=90}{h1}	
		\fmfv{decor.shape=triangle, decor.filled=full, decor.size=0.15w,d.a=105}{h2}
\end{fmfgraph*}}\quad=-~
	\raisebox{-0.75cm}{
	\begin{fmfgraph*}(60,48)
		\fmfpen{thin}
		\fmfleft{h1}
		\fmfright{h2,h3,h4}
		\fmfdot{a1}
		\fmf{photon,label=$\bullet$,l.d=0.05cm,tension=1}{a1,v1}
		\fmf{photon}{v1,h1}
		\fmf{photon,tension=0.5}{a1,h2}
		\fmf{photon,tension=0.5}{a1,h3}
		\fmf{photon,tension=0.5}{a1,h4}
		\fmfv{label=$a$}{h4}
		\fmfv{label=$c$,l.a=90}{h1}
		\fmfv{label=$b$,l.a=90}{h3}
		\fmfv{label=$d$}{h2}
		\fmffreeze
		\fmfsurroundn{j}{361}
		\fmfv{decor.shape=square,decor.filled=empty,decor.size=0.1w}{v1}
			\fmfv{decor.shape=triangle, decor.filled=full, decor.size=0.15w}{j110}
		\fmf{dbl_wiggly}{j110,v1}
\end{fmfgraph*}}\quad~~
-	\raisebox{-0.75cm}{
	\begin{fmfgraph*}(60,48)
		\fmfpen{thin}
		\fmfleft{h1}
		\fmfright{h2,h3,h4}
		\fmfdot{a1}
		\fmf{photon,label=$\bullet$,l.d=0.05cm,tension=1}{a1,v1}
		\fmf{photon}{v1,h1}
		\fmf{photon,tension=0.5}{a1,h2}
		\fmf{photon,tension=0.5}{a1,h3}
		\fmf{photon,tension=0.5}{a1,h4}
		\fmfv{label=$c$}{h4}
		\fmfv{label=$d$,l.a=90}{h1}
		\fmfv{label=$a$,l.a=90}{h3}
		\fmfv{label=$b$}{h2}
		\fmffreeze
		\fmfsurroundn{j}{361}
		\fmfv{decor.shape=square,decor.filled=empty,decor.size=0.1w}{v1}
					\fmfv{decor.shape=triangle, decor.filled=full, decor.size=0.15w}{j110}
		\fmf{dbl_wiggly}{j110,v1}
\end{fmfgraph*}}\quad
-
	\raisebox{-0.75cm}{
	\begin{fmfgraph*}(60,48)
		\fmfpen{thin}
		\fmfleft{h1}
		\fmfright{h2,h3,h4}
		\fmfdot{a1}
		\fmf{photon,label=$\bullet$,l.d=0.05cm,tension=1}{a1,v1}
		\fmf{photon}{v1,h1}
		\fmf{photon,tension=0.5}{a1,h2}
		\fmf{photon,tension=0.5}{a1,h3}
		\fmf{photon,tension=0.5}{a1,h4}
		\fmfv{label=$d$}{h4}
		\fmfv{label=$b$,l.a=90}{h1}
		\fmfv{label=$c$,l.a=90}{h3}
		\fmfv{label=$a$}{h2}
		\fmffreeze
		\fmfsurroundn{j}{361}
		\fmfv{decor.shape=square,decor.filled=empty,decor.size=0.1w}{v1}
					\fmfv{decor.shape=triangle, decor.filled=full, decor.size=0.15w}{j110}
		\fmf{dbl_wiggly}{j110,v1}
\end{fmfgraph*}}\quad
-
	\raisebox{-0.75cm}{
	\begin{fmfgraph*}(60,48)
		\fmfpen{thin}
		\fmfleft{h1}
		\fmfright{h2,h3,h4}
		\fmfdot{a1}
		\fmf{photon,label=$\bullet$,l.d=0.05cm,tension=1}{a1,v1}
		\fmf{photon}{v1,h1}
		\fmf{photon,tension=0.5}{a1,h2}
		\fmf{photon,tension=0.5}{a1,h3}
		\fmf{photon,tension=0.5}{a1,h4}
		\fmfv{label=$c$}{h4}
		\fmfv{label=$a$,l.a=90}{h1}
		\fmfv{label=$b$,l.a=90}{h3}
		\fmfv{label=$d$}{h2}
		\fmffreeze
		\fmfsurroundn{j}{361}
		\fmfv{decor.shape=square,decor.filled=empty,decor.size=0.1w}{v1}
					\fmfv{decor.shape=triangle, decor.filled=full, decor.size=0.15w}{j110}
		\fmf{dbl_wiggly}{j110,v1}
\end{fmfgraph*}}
}\\
	\bal{
	\raisebox{-0.75cm}{
		\begin{fmfgraph*}(60,48)
			\fmfpen{thin}
			\fmfleft{h1,h2}
			\fmfright{h3,h4}
			\fmfdot{a1}
			\fmf{photon,tension=0.5}{h3,a1}
			\fmf{dbl_wiggly,tension=0.5}{a1,h1}
			\fmf{fermion,tension=0.5}{h2,a1}
			\fmf{fermion,tension=0.5}{a1,h4}
			\fmfv{label=$i$}{h2}
			\fmfv{label=$j$}{h4}
			\fmfv{label=$a$}{h3}
				\fmfv{label=$a$}{h3}	\fmfv{decor.shape=triangle, decor.filled=full, decor.size=0.15w,d.a=135}{h1}
	\end{fmfgraph*}}\quad= 	
 - \raisebox{-0.75cm}{
\begin{fmfgraph*}(60,48)
\fmfpen{thin}
\fmfleft{h2}
\fmfright{h3,h4}
\fmfdot{a1}
\fmf{photon,tension=1}{h2,v1}
\fmf{photon,label=$\bullet$,l.d=0.05cm,tension=1}{a1,v1}
\fmf{fermion,tension=0.5}{h3,a1}
\fmf{fermion,tension=0.5}{a1,h4}
\fmfv{label=$\bar i$}{h3}
\fmfv{label=$j$}{h4}
\fmfv{label=$a$,l.a=90}{h2}
\fmffreeze
\fmfsurroundn{j}{360}
\fmf{dbl_wiggly}{v1,j117}
\fmfv{decor.shape=square,decor.filled=empty,decor.size=0.1w}{v1}
			\fmfv{decor.shape=triangle, decor.filled=full, decor.size=0.15w}{j117}
\end{fmfgraph*}}}\\
	\bal{
	\raisebox{-0.75cm}{
		\begin{fmfgraph*}(60,48)
			\fmfpen{thin}
			\fmfleft{h1,h2}
			\fmfright{h3,h4}
			\fmfdot{a1}
			\fmf{photon,tension=0.5}{h3,a1}
			\fmf{dbl_wiggly,tension=0.5}{a1,h1}
			\fmf{scalar,tension=0.5}{h2,a1}
			\fmf{scalar,tension=0.5}{a1,h4}
			\fmfv{label=$\bar i$}{h2}
			\fmfv{label=$j$}{h4}
			\fmfv{label=$a$}{h3}
				\fmfv{label=$a$}{h3}	\fmfv{decor.shape=triangle, decor.filled=full, decor.size=0.15w,d.a=135}{h1}
	\end{fmfgraph*}}\quad= 	\raisebox{-0.75cm}{
		\begin{fmfgraph*}(60,48)
			\fmfpen{thin}
			\fmfleft{h1,h2}
			\fmfright{h3,h4}
			\fmfv{decor.shape=circle,decor.filled=empty,decor.size=0.2w}{a1}
			\fmf{photon,tension=0.5}{h3,a1}
			\fmf{dbl_wiggly,tension=0.5}{a1,h1}
			\fmf{scalar,tension=0.5}{h2,a1}
			\fmf{scalar,tension=0.5}{a1,h4}
			\fmfv{label=$\bar i$}{h2}
			\fmfv{label=$j$}{h4}
			\fmfv{label=$a$}{h3}
				\fmfv{label=$a$}{h3}	\fmfv{decor.shape=triangle, decor.filled=full, decor.size=0.15w,d.a=135}{h1}
	\end{fmfgraph*}}\quad - \raisebox{-0.75cm}{
		\begin{fmfgraph*}(60,48)
			\fmfpen{thin}
			\fmfleft{h2}
			\fmfright{h3,h4}
			\fmfdot{a1}
			\fmf{photon,tension=1}{h2,v1}
			\fmf{photon,label=$\bullet$,l.d=0.05cm,tension=1}{a1,v1}
			\fmf{scalar,tension=0.5}{h3,a1}
			\fmf{scalar,tension=0.5}{a1,h4}
			\fmfv{label=$\bar i$}{h3}
			\fmfv{label=$j$}{h4}
			\fmfv{label=$a$,l.a=90}{h2}
			\fmffreeze
			\fmfsurroundn{j}{360}
			\fmf{dbl_wiggly}{v1,j117}
			\fmfv{decor.shape=square,decor.filled=empty,decor.size=0.1w}{v1}
						\fmfv{decor.shape=triangle, decor.filled=full, decor.size=0.15w}{j117}
\end{fmfgraph*}}}\\
	\bal{
	\raisebox{-0.75cm}{
		\begin{fmfgraph*}(60,48)
			\fmfpen{thin}
			\fmfsurround{h1,h2,h3,h4,h5}
			\fmfdot{a1}
			\fmf{photon,tension=0.5}{h2,a1}
			\fmf{photon,tension=0.5}{a1,h1}
			\fmf{scalar,tension=0.5}{a1,h3}
			\fmf{scalar,tension=0.5}{h4,a1}
			\fmf{dbl_wiggly,tension=0.5}{a1,h5}
			\fmfv{label=$\bar i$}{h4}
			\fmfv{label=$j$}{h3}
			\fmfv{label=$a$,l.a=90}{h1}
			\fmfv{label=$b$}{h2}
				\fmfv{label=$a$}{h3}	\fmfv{decor.shape=triangle, decor.filled=full, decor.size=0.15w,d.a=-145}{h5}
	\end{fmfgraph*}}
	\quad= - 	\raisebox{-0.75cm}{
		\begin{fmfgraph*}(60,48)
			\fmfpen{thin}
			\fmfleft{h1}
			\fmfright{h2,h3,h4}
			\fmfdot{a1}
			\fmf{photon,tension=0.3}{h2,a1}
			\fmf{photon,tension=1,label=$\bullet$,l.d=0.05cm}{a1,v1}
			\fmf{photon,tension=1}{v1,h1}
			\fmf{scalar,tension=0.3}{a1,h3}
			\fmf{scalar,tension=0.3}{h4,a1}
			\fmfv{label=$\bar i$}{h4}
			\fmfv{label=$j$}{h3}
			\fmfv{label=$a$,l.a=90}{h1}
			\fmfv{label=$b$}{h2}
			\fmffreeze
			\fmfsurroundn{j}{360}
			\fmfv{decor.shape=square,decor.filled=empty,decor.size=0.1w}{v1}
						\fmfv{decor.shape=triangle, decor.filled=full, decor.size=0.15w}{j115}
			\fmf{dbl_wiggly}{j115,v1}
\end{fmfgraph*}}
\quad~~- 	\raisebox{-0.75cm}{
	\begin{fmfgraph*}(60,48)
		\fmfpen{thin}
		\fmfleft{h1}
		\fmfright{h2,h3,h4}
		\fmfdot{a1}
		\fmf{photon,tension=0.3}{h4,a1}
		\fmf{photon,tension=1,label=$\bullet$,l.d=0.05cm}{a1,v1}
		\fmf{photon,tension=1}{v1,h1}
		\fmf{scalar,tension=0.3}{a1,h2}
		\fmf{scalar,tension=0.3}{h3,a1}
		\fmfv{label=$a$}{h4}
		\fmfv{label=$\bar i$}{h3}
		\fmfv{label=$b$,l.a=90}{h1}
		\fmfv{label=$j$}{h2}
		\fmffreeze
		\fmfsurroundn{j}{360}
		\fmfv{decor.shape=square,decor.filled=empty,decor.size=0.1w}{v1}
					\fmfv{decor.shape=triangle, decor.filled=full, decor.size=0.15w}{j115}
		\fmf{dbl_wiggly}{j115,v1}
\end{fmfgraph*}}}\\
with the identification of the effective Feynman rules\\
\begin{fleqn}
\bal{
\raisebox{-0.75cm}{
	\begin{fmfgraph*}(60,48)
		\fmfpen{thin}
		\fmfleft{h1,h2}
		\fmfright{h3,h4}
		\fmfv{decor.shape=circle, decor.filled=empty,decor.size= 0.2w}{a1}
		\fmf{dbl_wiggly,tension=0.5}{a1,h1}
		\fmf{photon,tension=0.5}{a1,h2}
		\fmf{photon,tension=0.5}{a1,h3}
		\fmf{photon,tension=0.5}{a1,h4}
		\fmfv{label=$b$}{h4}
		\fmfv{label=$c$}{h2}
		\fmfv{label=$a$}{h3}
			\fmfv{label=$a$}{h3}	\fmfv{decor.shape=triangle, decor.filled=full, decor.size=0.15w,d.a=135}{h1}
\end{fmfgraph*}} = \frac{\kappa g}{2} f^{abc}\Big[ \eta^{\mu_a\mu_b}\varepsilon_P^{\mu_c}(\varepsilon_P\cdot(p_b-p_a)) + \eta^{\mu_a\mu_c}\varepsilon_P^{\mu_b}(\varepsilon_P\cdot(p_a-p_c))+ \eta^{\mu_b\mu_c}\varepsilon_P^{\mu_a}(\varepsilon_P\cdot(p_c-p_b))\Big]
\label{eq:EffectiveVertexGGGGr}}
\\
\bal{
\raisebox{-0.75cm}{
	\begin{fmfgraph*}(60,48)
		\fmfpen{thin}
		\fmfleft{h1,h2}
		\fmfright{h3,h4}
		\fmfv{decor.shape=circle,decor.filled=empty,decor.size=0.2w}{a1}
		\fmf{photon,tension=0.5}{h3,a1}
		\fmf{dbl_wiggly,tension=0.5}{a1,h1}
		\fmf{scalar,tension=0.5}{h2,a1}
		\fmf{scalar,tension=0.5}{a1,h4}
		\fmfv{label=$\bar i$}{h2}
		\fmfv{label=$j$}{h4}
		\fmfv{label=$a$}{h3}
			\fmfv{label=$a$}{h3}	\fmfv{decor.shape=triangle, decor.filled=full, decor.size=0.15w,d.a=135}{h1}
\end{fmfgraph*}} = -\frac{i\kappa g}{2}\,\varepsilon_P^{\mu_a}(\varepsilon_P\cdot(p_i-p_j))\,(T^a)_{j\bar i}~.\label{eq:EffectiveVertexSSGGr}}
\end{fleqn}\\
The last two on-shell Feynman rules are\\\\\\\\
\begin{minipage}{0.5\textwidth}
	\bal{
	\raisebox{-0.9cm}{
		\begin{fmfgraph*}(60,60)
			\fmfpen{thin}
			\fmfsurroundn{h}{3}
			\fmfdot{a1}
			\fmf{dbl_wiggly,tension=0.5}{a1,h1}
			\fmf{quark,tension=0.3}{h2,a1}
			\fmf{quark,tension=0.3}{a1,h3}
			\fmfv{label=$\bar i$}{h2}
			\fmfv{label=$j$}{h3}
			\fmfv{decor.shape=triangle, decor.filled=full, decor.size=0.15w,d.a=-90}{h1}
	\end{fmfgraph*}}
	~~
	=-\frac{i\kappa}{2}\,\slashed{\varepsilon}_P\,(\varepsilon_P\cdot p_i)\delta^{j\bar i}}
\end{minipage}
\begin{minipage}{0.5\textwidth}
\bal{
	\raisebox{-0.9cm}{
		\begin{fmfgraph*}(60,60)
			\fmfpen{thin}
			\fmfsurroundn{h}{3}
			\fmfdot{a1}
			\fmf{dbl_wiggly,tension=0.5}{a1,h1}
			\fmf{scalar,tension=0.3}{h2,a1}
			\fmf{scalar,tension=0.3}{a1,h3}
			\fmfv{label=$\bar i$}{h2}
			\fmfv{label=$j$}{h3}
			\fmfv{decor.shape=triangle, decor.filled=full, decor.size=0.15w,d.a=-90}{h1}
\end{fmfgraph*}}~~=i\kappa\,(\varepsilon_P\cdot p_i)\,(\varepsilon_P\cdot p_j)\,\delta^{j\bar i}}
\end{minipage}
\\\\\\
which we can compare to the on-shell Feynman rules of pure YM - one of the gluon legs is taken to be external and contracted with its polarization tensor. Consider\\\\\\\\
\begin{minipage}{0.5\textwidth}
	\bal{
		\raisebox{-0.9cm}{
			\begin{fmfgraph*}(60,60)
				\fmfpen{thin}
				\fmfsurroundn{h}{3}
				\fmfdot{a1}
				\fmf{photon,tension=0.5}{a1,h1}
				\fmf{quark,tension=0.3}{h2,a1}
				\fmf{quark,tension=0.3}{a1,h3}
				\fmfv{label=$\bar i$}{h2}
				\fmfv{label=$j$}{h3}
				\fmfv{label=$p$}{h1}
				\fmfv{decor.shape=triangle, decor.filled=full, decor.size=0.15w,d.a=-90}{h1}
		\end{fmfgraph*}}
		~~~~
		=ig\,\slashed{\varepsilon}_p\,(T^p)_{j\bar i}}
\end{minipage}
\begin{minipage}{0.5\textwidth}
	\bal{
		\raisebox{-0.9cm}{
			\begin{fmfgraph*}(60,60)
				\fmfpen{thin}
				\fmfsurroundn{h}{3}
				\fmfdot{a1}
				\fmf{photon,tension=0.5}{a1,h1}
				\fmf{scalar,tension=0.3}{h2,a1}
				\fmf{scalar,tension=0.3}{a1,h3}
				\fmfv{label=$\bar i$}{h2}
				\fmfv{label=$j$}{h3}
				\fmfv{label=$p$}{h1}
				\fmfv{decor.shape=triangle, decor.filled=full, decor.size=0.15w,d.a=-90}{h1}
		\end{fmfgraph*}}~~~~=2ig\,(\varepsilon_p\cdot p_i)\,(T^p)_{j\bar i}~~}
\end{minipage}
\\\\\\\\
where we used momentum conservation in the two scalar one gluon Feynman rule $p_j = -p_i - p_a$. Upon a closer look we notice that we can reach the effective vertices with an external graviton via the prescription
\bal{
(T^p)_{j\bar i} \rightarrow \frac{\kappa}{2g}(\varepsilon_p\,p_j)\,\delta^{j\bar i}\quad,\quad (T^p)_{j\bar i} \rightarrow -\frac{\kappa}{2g}(\varepsilon_p\,p_i)\,\delta^{j\bar i}~.
}
 Similarly we obtain\\
\bal{
	\raisebox{-0.75cm}{
		\begin{fmfgraph*}(60,48)
			\fmfpen{thin}
			\fmfsurroundn{h}{3}
			\fmfdot{a1}
			\fmf{photon,tension=0.5}{a1,h1}
			\fmf{photon,tension=0.5}{a1,h2}
			\fmf{photon,tension=0.5}{a1,h3}
			\fmfv{label=$b$}{h3}
			\fmfv{label=$p$}{h1}
			\fmfv{label=$a$}{h2}
			\fmfv{decor.shape=triangle, decor.filled=full, decor.size=0.15w,d.a=-90}{h1}
	\end{fmfgraph*}}
	\quad~~
	=2gf^{abp}\left[M^{\mu_a\mu_b} + \frac{1}{2}\,W^{\mu_a\mu_b}\right]~.}\\\\
Ignoring the Ward identity term, we can generate the effective vertex with one external graviton \eqref{eq:effectiveVertexGGGr} through
\bal{
	f^{apb} = i(T^p)_{ab} \rightarrow -i\frac{\kappa}{2g}(\varepsilon_p\,p_a)\,\delta^{ab}\quad,\quad i(T^p)_{ab} \rightarrow i\frac{\kappa}{2g}(\varepsilon_p\,p_b)\,\delta^{ab}~.
} 
We observe that the prescription of the fundamental color factors differs to the adjoint factors by a minus sign. Nevertheless, we can define adjoint color factors $\tilde f^{abc} = i f^{abc}$ which will have the exact same behaviour in the color to kinematics prescription. Higher multiplicity vertices can also be generated by this prescription. We start with the case\\
	\bal{
	\raisebox{-0.9cm}{
		\begin{fmfgraph*}(60,48)
			\fmfpen{thin}
			\fmfleft{h1,h2}
			\fmfright{h3,h4}
			\fmfdot{a1}
			\fmf{photon,tension=0.5}{h3,a1}
			\fmf{photon,tension=0.5}{a1,h1}
			\fmf{scalar,tension=0.5}{h2,a1}
			\fmf{scalar,tension=0.5}{a1,h4}
			\fmfv{label=$\bar i$}{h2}
			\fmfv{label=$j$}{h4}
			\fmfv{label=$a$}{h1}
			\fmfv{label=$p$}{h3}
			\fmfv{decor.shape=triangle, decor.filled=full, decor.size=0.15w,d.a=-135}{h3}
	\end{fmfgraph*}}
	\quad~~
	=ig^2\varepsilon^{\mu_a}_p\left((T^a)_{jk}(T^p)_{k\bar i}+(T^p)_{jk}(T^a)_{k\bar i}\right)}
\\
Sending 
\bal{
(T^p)_{k\bar i} \rightarrow = -\frac{\kappa}{2g}(\varepsilon_p\cdot p_i)\,\delta^{k\bar i}\quad,\quad (T^p)_{jk} \rightarrow \frac{\kappa}{2g}(\varepsilon_p\cdot p_j)\,\delta^{jk}~,
}
we can see that we land exactly on \eqref{eq:EffectiveVertexSSGGr}. The four gluon vertex is more involved. We start by writing \\
	\bal{
	\raisebox{-0.75cm}{
		\begin{fmfgraph*}(60,48)
			\fmfpen{thin}
			\fmfleft{h1,h2}
			\fmfright{h3,h4}
			\fmfdot{a1}
			\fmf{photon,tension=0.5}{a1,h1}
			\fmf{photon,tension=0.5}{a1,h2}
			\fmf{photon,tension=0.5}{a1,h3}
			\fmf{photon,tension=0.5}{a1,h4}
			\fmfv{label=$b$}{h4}
			\fmfv{label=$p$}{h1}
			\fmfv{label=$a$}{h3}
			\fmfv{label=$c$}{h2}
			\fmfv{decor.shape=triangle, decor.filled=full, decor.size=0.15w,d.a=140}{h1}
	\end{fmfgraph*}}~~=-ig^2\Big[&\varepsilon^{\mu_c}_p\,\eta^{\mu_b\mu_a}(f^{cbe}f^{pae}-f^{cae}f^{bpe}) - \eta^{\mu_c\mu_a}\,\varepsilon^{\mu_b}_p\,(f^{cbe}f^{pae}+f^{cpe}f^{bae}) \\[-\baselineskip]&+ \eta^{\mu_c\mu_b}\,\varepsilon^{\mu_a}_p(f^{cpe}f^{bae}+f^{cae}f^{bpe})\Big]~.}\\
Let us focus on the first color structure
\bal{f^{cbe}f^{pae}-f^{cae}f^{bpe} &= -(f^{cbe}f^{ape}+f^{cae}f^{bpe})= -i(f^{cbe}\,(T^p)_{ae}+f^{cae}(T^p)_{be}) \\& \rightarrow  i\frac{\kappa}{2g}\left(f^{cbe}\,(\varepsilon_p\cdot p_a)\,\delta^{ae}+f^{cae}(\varepsilon_p\cdot p_b)\right)\delta^{be} = i\frac{\kappa}{2g} \left(f^{cba}\,(\varepsilon_p\cdot p_a)+f^{cab}(\varepsilon_p\cdot p_b)\right)\\& = i\frac{\kappa}{2g} f^{abc}\,(\varepsilon_p\cdot (p_b-p_a))~.}
Hence
\bal{
-ig^2\Big[\varepsilon^{\mu_c}_p\,\eta^{\mu_b\mu_a}(f^{cbe}f^{pae}-f^{cae}f^{bpe})\Big] \rightarrow \frac{g\kappa}{2}\,\varepsilon^{\mu_c}_p\,\eta^{\mu_b\mu_a}\, f^{abc}\,(\varepsilon_p\cdot (p_b-p_a))~,}
which is exactly the first term in \eqref{eq:EffectiveVertexGGGGr}. The other terms are reached in exactly the same fashion. This section concludes two statements
\begin{enumerate}
	\item From the notion of effective on-shell Feynman rules we can deduce that the number of possible graviton insertions into a tree level diagram is the same as the amount of possible gluon insertions into the same diagram.
	\item The effective on-shell Feynman rules with one external graviton, and therefore diagrams including such vertices, can be generated by the proposed double copy from pure gluon Feynman rules.
\end{enumerate} 
\section{Proof of Key Relation}
Eq. \eqref{CC-replacement} has already been proven in the case of $\sigma_b=c$ being a gluon. It is straightforward to prove the other cases.
\begin{enumerate}
	\item \textbf{$\sigma_b=c$ is a anti-quark:} The color factor will have some expression that depends on the Dyck word. Nevertheless we may zoom in on the position of insertion and read
	\be
	C_{\ldots|cp|\ldots}= \cdots\{c|\,T^c\,\otimes\, \Xi_{l-1}^c\,\Xi^p_{l} \,\cdots=\cdots\{c|\,(T^c\,T^p)\,\otimes\, \Xi_{l-1}^c \,\cdots + \cdots\{c|\,(T^c\,\otimes\, (\Xi_{l-1}^c\,\Xi^p_{l-1}) \,\cdots~,
	\ee
	where we used the definition \eqref{xidef}. Similarly we get
	\be
	C_{\ldots|pc|\ldots}= \cdots\Xi^p_{l-1}\,\{c|\,T^c\,\otimes\, \Xi_{l-1}^c \,\cdots= \cdots\{c|\,(T^c\,\otimes\, (\Xi_{l-1}^p\,\Xi^c_{l-1}) \,\cdots~.
	\ee
	Now computing the commutator will give
	\bal{
		C_{\ldots|cp|\ldots} &- C_{\ldots|pc|\ldots}  = \cdots\{c|\,(T^c\,T^p)\,\otimes\, \Xi_{l-1}^c \,\cdots + \cdots\{c|\,T^c\,\otimes\,\left(\left[\Xi_{l-1}^c\,,\,\Xi^p_{l-1}\right]\right))\cdots\\
		&=\cdots\{c|\,(T^p\,T^c)\,\otimes\, \Xi_{l-1}^c \,\cdots\,+\,\cdots\{c|\,\left(\left[T^c\,,\,T^p\right]\right)\,\otimes\, \Xi_{l-1}^c \,\cdots + \cdots\{c|\,T^c\,\otimes\,\left(\left[\Xi_{l-1}^c\,,\,\Xi^p_{l-1}\right]\right))\cdots
		\\&=\cdots\{c|\,(T^p\,T^c)\,\otimes\, \Xi_{l-1}^c \,\cdots\,+\,\cdots\,\tilde f^{cpk}\{c|\,\left(T^k\right)\,\otimes\, \Xi_{l-1}^c \,\cdots + \cdots\,\tilde f^{cpk}\{c|\,(T^c)\,\otimes\,\Xi_{l-1}^k\cdots
		\\&=\cdots\{c|\,(T^p\,T^c)\,\otimes\, \Xi_{l-1}^c \,\cdots~,}
	immediately implying
	\be
	\left.\left(C_{\ldots|cp|\ldots}-C_{\ldots|pc|\ldots~}\right)\right|_{\mathcal R_p} = (\varepsilon_{p}\cdot k_{c}) \, C_{\ldots|c|\ldots}~.	
	\ee
	\item \textbf{$\sigma_b=c$ is a quark:} We zoom in and see
	\bal{
		C_{\ldots|cp|\ldots}&= \cdots |c\} \Xi^p_{l} \cdots~, \\C_{\ldots|pc|\ldots}&= \cdots \Xi^p_{l+1} |c\}  \cdots = \cdots T^p |c\} \cdots + \cdots |c\} \Xi^p_{l} \cdots~.
	}
Hence the commutator is just
\bal{
C_{\ldots|cp|\ldots} - C_{\ldots|pc|\ldots} = - \cdots T^p |c\} \cdots~,
} 
and we immediately obtain 
\be
\left.\left(C_{\ldots|cp|\ldots} - C_{\ldots|pc|\ldots}\right)\right|_{\mathcal R_p} = (\varepsilon_{p}\cdot k_{c}) \, C_{\ldots|c|\ldots}~.
\ee
This concludes the proof of \eqref{CC-replacement}.
\end{enumerate}
\end{fmffile}

\end{document}